\documentclass[fleqn,usenatbib]{mnras}

\usepackage[T1]{fontenc}

\DeclareRobustCommand{\VAN}[3]{#2}
\let\VANthebibliography\thebibliography
\def\thebibliography{\DeclareRobustCommand{\VAN}[3]{##3}\VANthebibliography}

\usepackage{floatrow}
\usepackage{graphicx}	
\usepackage{amsmath}	
\usepackage{amssymb}	
 
\usepackage{soul}
\soulregister\cite7
\soulregister\citep7
\soulregister\citealp7
\soulregister\citet7
\usepackage{booktabs}
\usepackage{textgreek}
\usepackage{tikz}
\usepackage{appendix}
\usepackage{enumerate}
\usepackage{subcaption}
\usepackage{hyperref}
\usepackage{cleveref}

\usepackage{newtxtext,newtxmath}

\newcommand{\ntild}{\char`\~}
\newcommand\abs[1]{\left|#1\right|}
\newcommand\embrace[1]{\{#1\}}

\title[Machines Learn to Infer Stellar Parameters]{Machines Learn to Infer Stellar Parameters Just by Looking \\at a Large Number of Spectra}

\author[N. Sedaghat et al.]{
Nima Sedaghat,$^{1}$\thanks{E-mail: nimaseda@uw.edu}
Martino Romaniello,$^{1}$
Jonathan E. Carrick$^{2}$
and François-Xavier Pineau$^{3}$
\\
$^{1}$European Southern Observatory (ESO), Karl-Schwarzschild-Str., 85748 Garching, Germany\\
$^{2}$Physics Department, Lancaster University, Bailrigg, Lancaster, LA1 4YW, UK\\
$^{3}$Université de Strasbourg, CNRS, Observatoire astronomique de Strasbourg, UMR 7550, F-67000 Strasbourg, France
}

\date{Accepted XXX. Received YYY; in original form ZZZ}

\pubyear{2020\textbf{}}

\begin{document}
\label{firstpage}
\pagerange{\pageref{firstpage}--\pageref{lastpage}}
\maketitle

\begin{abstract}
Machine learning has been widely applied to clearly defined problems of astronomy and astrophysics. However, deep learning and its conceptual differences to classical machine learning have been largely overlooked in these fields.
The broad hypothesis behind our work is that letting the abundant real astrophysical data speak for itself, with minimal supervision and no labels, can reveal interesting patterns which may facilitate discovery of novel physical relationships.
Here as the first step, we seek to interpret the representations a deep convolutional neural network chooses to learn, and find correlations in them with current physical understanding. 
We train an encoder-decoder architecture on the self-supervised auxiliary task of reconstruction to allow it to learn general representations without bias towards any specific task. By exerting weak disentanglement at the \textit{information bottleneck} of the network, we implicitly enforce interpretability in the learned features. We develop two independent statistical and information-theoretical methods for finding the number of learned \textit{informative features}, as well as measuring their true correlation with astrophysical validation labels.
As a case study, we apply this method to a dataset of $\sim$270000 stellar spectra, each of which comprising $\sim$300000 dimensions. We find that the network clearly assigns specific nodes to estimate (notions of) parameters such as radial velocity and effective temperature without being asked to do so, all in a completely physics-agnostic process. This supports the first part of our hypothesis.
Moreover, we find with high confidence that there are $\sim$4 more independently informative dimensions that do not show a direct correlation with our validation parameters, presenting  potential room for future studies.

\end{abstract}


\section{Introduction}

Big Data has already changed the way we do science in nearly all areas of research everyday. Although \textit{data-driven} methods have been around since almost the very beginning of the history of science, the meaning of the term has started to transform gradually; data is not used only to validate our analytical formulations and hypotheses any more, but has started taking more serious roles in defining the problem itself, and providing non-parametric solutions to it.

\begin{figure}
  \begin{center}
    \includegraphics[width=\linewidth]{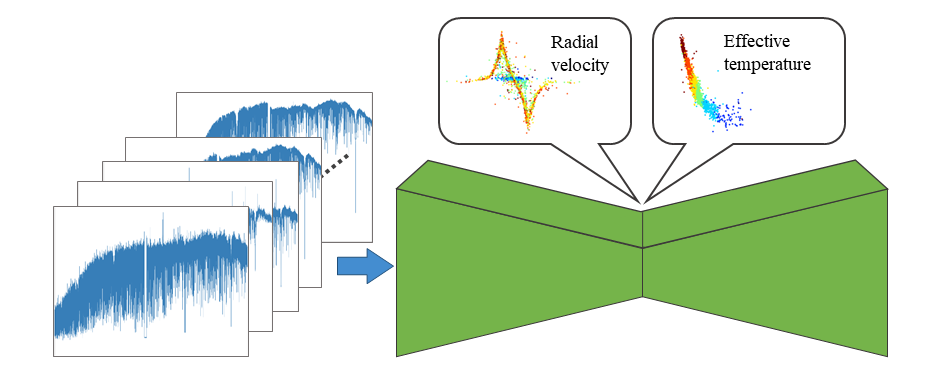}
  \end{center}
   \caption{
   A large number of stellar spectra are passed through the    \textit{information bottleneck} of a deep convolutional autoencoder, in a fully unsupervised, physics-agnostic process. The network has zero information about the content of the numerical vectors it receives. We use techniques based on information maximization, to enforce learning of disentangled features, and find that the network learns representations for astrophysical parameters such as \textit{radial velocity} and \textit{effective temperature}, without being asked to do so.}
  \label{fig:teaser}
\end{figure}

The rationale behind this reform is two-fold. First, huge amounts of new data are becoming available in many areas: from the ever increasing number of search-able images on the web to the petabytes-per-minute streams of data expected from future telescopes -- e.g., see SKA \citep[][]{quinn2015delivering}. 
Secondly, and perhaps more importantly, the scientific community has found, and is advancing, ways to handle such big volumes of data, thanks to advances in technology. At the core of these advances lies the recent revolution of techniques under the broad term of \textit{machine learning}.

The number of machine learning-based solutions to problems in astrophysics, astronomy and cosmology has drastically increased in the past years, and providing a list of them is beyond the scope of this manuscript -- we refer to \citet{baron2019machine} for a practical overview. We believe what particularly needs to be assessed, however, is the way learning has been utilized in these fields, and the potentials to broaden the horizons.
Concretely speaking, the so-called revolution of the past two decades has been more about \textit{deep learning}~\citep[DL,][]{krizhevsky_imagenet_2012,raina_large-scale_2009}: a new family of methods \textit{forked} out of classical machine learning (ML) -- the latter has already been around since as early as 1980's~\citep{lecun_procedure_1985}.
But most of the solutions used by our community have been plugin-style usages of classical ML, and the advantages deep learning brings upon have not found enough exposure.

Classical ML can be roughly modeled as a black box which implicitly learns how to connect input features (engineered by humans) to desired output. Deep learning, on the other hand, is a similar box, normally implemented as a neural network, with the additional capability to learn and decide what features are best to be used for the task at hand. The ability, also known as \textit{representation learning}~\citep{rumelhart_1986_learning,bengio_representation_2014}, is the key difference between the two methodologies -- not the depth of the neural network.

Nevertheless, deep models have proven superiority in performance and accuracy over traditional methods in astronomy and astrophysics. Applications involving classification, detection, regression, have been extensively and successfully outsourced to neural networks in the past years; from redshift estimation~\citep{vanzella_photometric_2004} to morphological classification~\citep{lukic_galaxy_2016}. Yet, there has been little work towards finding \textit{how} a network is tackling a specific problem and indeed the interpretation of what the network has learned is still an open line of research, in all areas.

Unsupervised approaches have also been extensively studied, especially in the field of computer vision where deep learning was originally cultivated; E.g. see \citet{bengio_unsupervised_2012} for a review. Such methods have even been attempted in other fields of science too, including astronomy (e.g. see \citealp{baron2017weirdest}). However, they have often been used to either learn proper features for initialization of the \textit{main} supervised task (e.g. \citealp{martinazzo2020self}), or simply as techniques for tasks such as dimensionality reduction~\citep{hinton_reducing_2006}, compression~\citep{wulff2020deep}, storage tractability. 

In this work, we choose to take a fully unsupervised approach, without defining any specific tasks for the network. The idea is to attempt to interpret the representations by which the network decides to perceive and describe the data, and assess whether there are traces of (astro-)physical concepts in them.

The idea of ``distilling data into knowledge'' in form of analytical expressions was introduced by \citealp[][]{schmidt2009distilling}, and later adapted to astronomy \citep{graham2013machine} and cosmology \cite{krone2014first}. 
Our work shares the same basic goal at the conceptual level: letting a machine learn from experimental data. However, we go beyond the constraints of analytical expressions and try to capture the knowledge in a non-parametric fashion, relying on the hierarchical feature learning capabilities of deep neural networks.

In the past years, there have been works lying at the cross-section of deep learning and the broad definition of the term \textit{physics}. Most of such works implement \textit{physics-guided} or \textit{physics-informed} networks, where the network is, explicitly or implicitly pre-fed with known physical laws (e.g. see \citealp{zhang_physics-guided_2020,meng_ppinn_2020}). Inspired by Hamiltonian mechanics, \citet{greydanus_hamiltonian_2019, choudhary_physics-enhanced_2020} design Hamiltonian Neural Networks that learn to respect exact conservation laws.
\citet{raissi_physics_2017} teach neural networks to solve tasks while respecting physical laws described by partial differential equations. 
\citet{stewart_label-free_2017} use prior knowledge to limit the space of possible learned mappings.
\citet{denil_learning_2017} use reinforcement learning to pursue physical experiments. 
\citet{ehrhardt_learning_2017} use simulated motion sequences to teach a neural network to predict motion, where \citet{sedaghat_hybrid_2017} predict motion patterns in real videos.
\citet{dagnolo_learning_2019,de_simone_guiding_2019} use neural networks to detect discrepancy between reference models and actual (synthetic) data.
However in all of them the flow of physics knowledge is, directly or indirectly, from human mind to the machine, whereas in this work, we focus on observing how the machines learn; i.e. the way Big Data enforces the machine to interpret it.

\Cref{fig:teaser} outlines our implementation of the above idea. We use an archive of stellar spectra obtained using the HARPS (High Accuracy Radial-velocity Planet Searcher, \citealp{pepe2002harps,santos2004harps,2018SPIE10704E..16R}) instrument, as an exemplar case for study, with easy access to a large number of samples~\footnote{We henceforth refer to the dataset itself as HARPS}. We pass the data, as a set of 1-D~\footnote{The term 1-D here is used the way it is used in the signal processing literature, to differentiate vectors from 2-D arrays, a.k.a matrices, and higher dimensionalities. Otherwise, from a computer scientific point of view each spectrum in our case has a dimensionality of \ntild 300000.} numerical arrays, through the \textit{information bottleneck} \footnote{We use the term `information bottleneck' in a loose manner for both the exact theory of \citet{tishby_information_2000}, as well as the architectural bottleneck formed where the encoder and decoder of an autoencoder meet.} of a deep convolutional autoencoder, seeking a low-dimensional yet informative representation of the data \citep{tishby_information_2000}. The process is fully unsupervised and the network is completely agnostic of the type of the content it is seeing. The only constraint we apply during training is enforcing disentanglement of the learned representations \citep{bengio_representation_2014}, based on maximization of the mutual information \citep{cover_elements_1991} between latent representations and the main signal. This, however, is the key component of our implementation, as we need to tune the disentanglement weight to a lower-than-standard level, for the method to work.

We crack open the trained network, and surprisingly find that clear traces of physical concepts, such as the effective temperature of stars and radial velocity are captured by the network. In other words, the network learns to identify and map such physical features to individual dedicated latent nodes. Such correlations are identified by seeking mutual information between the latent nodes and astrophysical validation labels we manage to collect from published catalogues (through the VizieR interface, \citealp{ochsenbein2000vizier}), for a subset of HARPS object.

In parallel, we define a purely statistical informativeness measure and run it on the latent nodes to find probable candidates for analysis. Although the weight we put on disentanglement affects the results, we find in a reasonable setting that 6 nodes (out of 128) supposedly capture a noticeable amount of information. Interestingly, the two \textit{physical nodes} we already identified are among the 6, leaving the remaining 4 open for future studies. As scientifically surprising as the identified physical nodes are, the remaining 4 are potentially even more important in the context of the long-term goal of our studies, as they may open doors for us to learn \textit{new} patterns/correlations from data.

Our implementation is based on autoencoders~\citep{vincent2010stacked}: the de-facto framework for unsupervised approaches in deep learning. The image generating capability of convolutional encoder-decoder architectures has also been utilized in for tasks such as transient detection~\citep{sedaghat2018effective} and de-blending~\citep{boucaud2020photometry}. However, we move from the deterministic version to Variational AutoEncoders (VAE: \citealp{kingma_auto-encoding_2014}), where statistical analysis is made possible. VAEs and their extensions have been widely used to achieve (or enforce) interpretability in latent representations -- e.g. see \citet{bengio_representation_2014, higgins2016beta, chen_isolating_2018, zhao_infovae_2018,tschannen_recent_2018,crescimanna_variational_2019}. A comprehensive tutorial on VAEs can be found in \citet{doersch_tutorial_2016}. Information-theoretic extensions to VAEs have also been studied recently by e.g. \citet{crescimanna_variational_2019} and \citet{rezaabad_learning_2020}.

\begin{figure*}[t]
  \begin{center}
    \includegraphics[width=\linewidth]{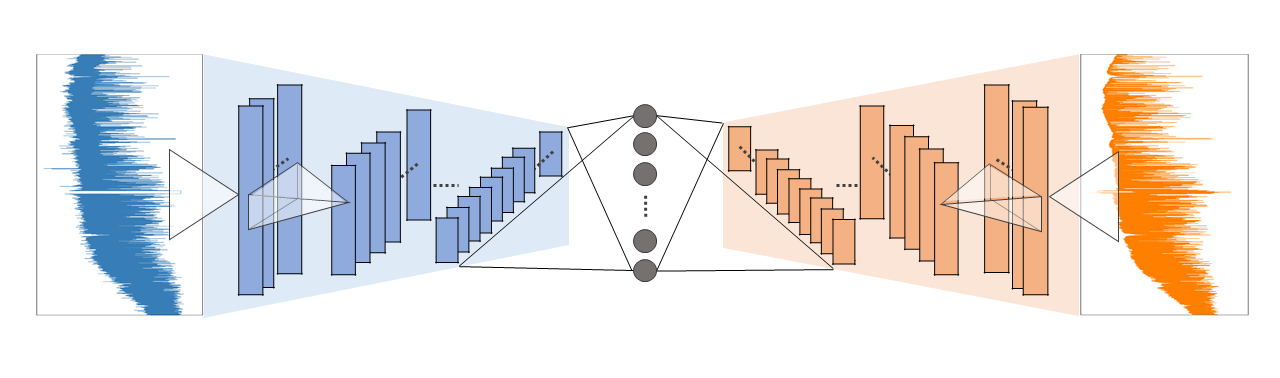}
    \\(a)

    \includegraphics[width=0.35\linewidth]{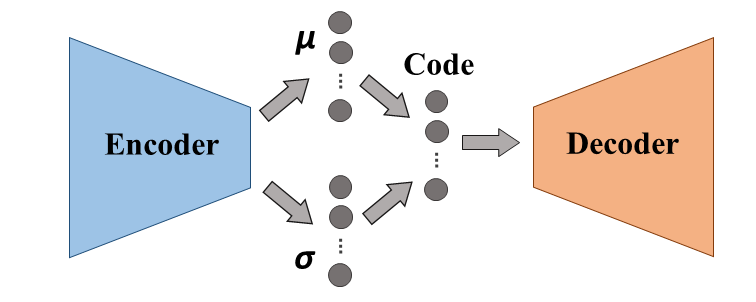}
    \\(b)
  \end{center}
\caption{Brief architecture of the deterministic autoencoder on top, with the schematic variational counterpart of it at the bottom. In the VAE version, the code is not directly connected to the encoder, but is drawn from the learnable parameters of the normal distribution: \textit{reparametrization trick} \citep{kingma_auto-encoding_2014}.}
  \label{fig:architecture}
\end{figure*}

Perhaps the closest to our implementation, is the parallel work of \citet{iten_discovering_2020} where a $\beta$-VAE is used to look for traces of physics in latent representations. However, in that work only ``toy examples'' based on simulations are tried, with a rather shallow non-convolutional network. This makes the work orthogonal to our long-term goal of ``learning from data'': simulations are created based on simplistic mathematical models we already know. Hence they can teach us, at best, the things we already know.

Moreover, for a network to be able to learn \textit{semantics} from data, it needs to be a) presented with huge amounts of \textit{real} data, to avoid overfitting and falling in the covariate shift trap \citep{sugiyama2012machine}, and b) at the same time sophisticated and deep enough to learn \textit{useful} representations.

There have also been a few attempts towards finding physical parameters in spectra based on typical dimensionality reduction methods such as Principal Component Analysis (PCA, \citet{jolliffe2016principal}). However, PCA provides a linear decomposition of data and hence, as expected, does not yield the desired one-to-one mapping between the principal components and physical features -- \citealp[e.g. see][]{bailer1998automated}. We illustrate such an effect on our dataset in \cref{app:pca}.

\paragraph*{Our Contributions}
\begin{itemize}
    \item To the best of our knowledge, this is the first work to allow deep convolutional neural networks to learn to infer (astro-)physical parameters just by looking at \textit{real} data, with \textit{zero supervision}.
    \item We provide methods based on mutual information and statistics, to track true correlation between learned representations and physical parameters, as well as auto-discovery of the potentially informative latent dimensions.
    \item We identify but leave open, cues for doing science with potentially new patterns that neural networks discover in data.
\end{itemize}

\noindent \Cref{sec:deterministic} presents the basic deterministic convolutional autoencoder we start our study with. \Cref{sec:enforce} explains how we enforce interpretability of the learned representations via disentanglement. \Cref{sec:dataset} details the specifications of the dataset. In \cref{sec:reconst} we briefly look at reconstruction results. Finally in \cref{sec:physics} we analyse the learned latent representations and assess traces of physics in them.

\section{A Deterministic Convolutional AutoEncoder}
\label{sec:deterministic}

Although the final implementation of the proposed method involves treatment of the input and the latent representation as statistical variables, in this section, we start by detailing the architecture of a \textit{deterministic} deep convolutional autoencoder~\citep{vincent2010stacked} and training details. 
This allows us to clarify the migration from a traditional \textit{fully-connected} autoencoder to a convolutional one, as well as to briefly illustrate that even the deterministic variant is capable of learning useful information from Big Data.

\paragraph*{Architecture}
We design an autoencoder composed of a combination of convolutional, up-convolutional and fully connected layers -- \Cref{fig:architecture}. A fully detailed illustration of the network architecture is presented in \cref{app:architecture}.
There are 15 convolutional layers in the \textit{encoder} part that transform the input spectrum, $x$, down to 512 vectors of length 20 (in case of HARPS). 
The vectors are then transformed to a single vector of scalars, called \textit{code}, using a fully connected layer. The code, also referred to as the \textit{latent representation} throughout this article, contains the most compressed version of the input spectrum throughout the network.
The dimensionality of this vector is chosen based on the the desired compression rate. We experiment with different code sizes, from 2 to 128.
On the other side of the bottleneck, a second fully-connected layer transforms the code back to a similar set of 512 vectors. Then a set of up-convolutional layers take them step-by-step up to the same dimensionality as the original input ($327680$ for HARPS).

\paragraph*{Reconstruction Loss}
$E_{\phi}$ and $D_{\theta}$ represent the deterministic encoder and decoder, respectively, where $\phi$ , $\theta$ are the learn-able parameters of the network.
We aim for pixel-level accuracy in the reconstructed spectrum and so choose to minimize the per-pixel L1 loss function:
\begin{equation}
\mathcal{L}_{AE}(\theta,\phi)=\mathbb{E}_{data}[||x-D_{\theta}(E_{\phi}(x))||^1_1]
\end{equation}

\noindent which is empirically computed as:

\begin{equation}
    \mathcal{L}_{AE}=\frac{\sum_{i\in\mathcal{M}}{\abs{x_i-\hat{x}_i}}}{n}
\end{equation}

\noindent where $x$, $\hat{x}$ are the input and reconstructed spectra respectively, $i$ is the pixel index and $n$ is the total number of pixels.

Set $\mathcal{M}$ represents a mask, constant over all the spectra in the dataset, which masks out the three \textit{information gaps} in the beginning, middle and end of each HARPS spectrum \citep{pepe2002harps}. This is a safe procedure, because these are just instrumental artifacts that bear no meaning for the astrophysical interpretation of the spectra. \footnote{The location of such artifacts is not exactly fixed across different spectra. Therefore we chose to use a single constant mask to cover all of them, at the cost of losing a small fraction of informative pixels from each spectrum.}

\paragraph*{Median Normalization}
For stability of the training process, we want the input samples not to feed extremely different value ranges into the input of the network. Thus, without loss of generality, we normalize the spectra in the dataset according to
\begin{equation}
    x = \dfrac{\mathring{x}}{\underset{i\in\mathcal{M}}{\mathrm{median}}{\{\mathring{x}_i\}}}
\end{equation}

\noindent in which $\mathring{x}$ is the original input spectrum before normalization.

Our initial experiments show that a deterministic autoencoder not only can compress and reconstruct the whole datasets with as few as 8 nodes at the bottleneck and with a high quality, but also can grasp a degree of understanding about the underlying signal sources. This is reflected in the way the network treats the telluric lines differently to other (stellar) lines. Details of this part of the study will be published in a future article.

\section{Enforcing Interpretability}
\label{sec:enforce}
Learning disentangled representations for composing factors of observed phenomena is key to interpretability \citep{bengio_representation_2014}.

Although our deterministic autoencoder proves to be capable of learning interesting aspects of the observations, the de-facto methods of enforcing disentanglement in deep autoencoders are built on top of the VAE-based family of methods, and are done by regularization of the variational autoencoder objective, one way or another \citep{tschannen_recent_2018}.

We convert our classic autoencoder to a VAE, as seen in \cref{fig:architecture}, where the deterministic code is replaced by a probabilistic one and each element of it is drawn from a normal distribution defined by a pair of learnable parameters: mean ($\mu$) and standard deviation ($\sigma$).

In the most basic form of a VAE, the objective is of the form:
\begin{equation}
    \mathcal{L}_{\scriptscriptstyle VAE}(\theta,\phi)=\mathcal{L}_{reconst}(\theta,\phi) + \mathbb{E}_{data}[D_{KL}(q_{\phi}(z|x)||p_{\theta}(z))]
\end{equation}

\noindent where $z$ is the latent variable, $p_{\theta}(z)$ is the prior distribution on the latent space. $q_{\phi}(z|x)$ is the approximation of the posterior, learned by the encoder and $D_{KL}$ represents the Kullback-Liebler divergence \citep{kullback1951information}.

\citet{higgins2016beta} introduce $\beta$-VAE in which more disentanglement is enforced by increasing the weight ($\lambda$) of the second term:

\begin{equation}
\label{eq:beta-vae-loss}
    \mathcal{L}(\theta,\phi)=\mathcal{L}_{reconst}(\theta,\phi) + \lambda \mathbb{E}_{data}[D_{KL}(q_{\phi}(z|x)||p_{\theta}(z))]
\end{equation}

\noindent which from another perspective, pushes for maximizing the mutual information between $z$ and $x$ -- \citealp[e.g. see][]{burgess_understanding_2018}. We follow the same formulation for enforcing disentanglement in our implementation. However, we find that pushing for too much disentanglement by setting $\lambda$ to too high a value, even values close to 1 as suggested by \citet{kingma_auto-encoding_2014,higgins2016beta}, results in too much loss of reconstruction quality, rendering it against the main goal of this work. We assess this trade-off between disentanglement and reconstruction quality in the upcoming sections and find $\lambda=0.3$ a reasonable choice for the current task.

\begin{figure*}
\begin{tabular}{ c c c c c }
    \multicolumn{1}{c}{\tiny{Deterministic AE}} & \multicolumn{1}{c}{\tiny{Disentangled VAE}} 
    &&
    \multicolumn{1}{c}{\tiny{Deterministic AE}} & \multicolumn{1}{c}{\tiny{Disentangled VAE}}
    \bigskip
    \\
    \includegraphics[width=.2\linewidth,height=0.2\linewidth]{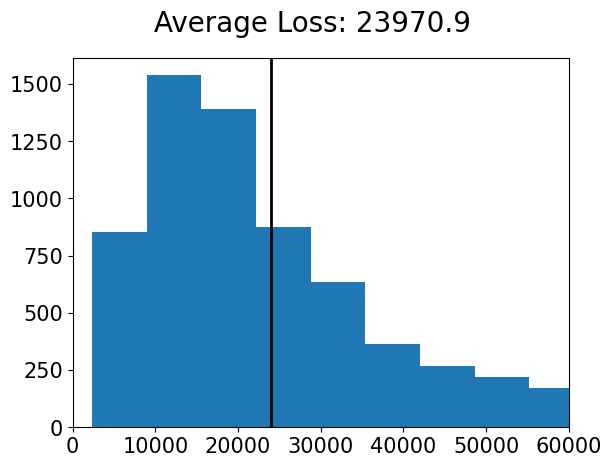}
    & 
    \includegraphics[width=.2\linewidth,height=0.2\linewidth]{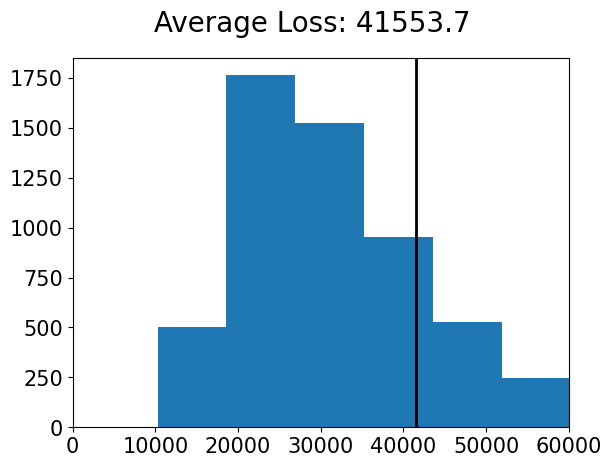}  
    &
    2-D
    &
    \includegraphics[width=.2\linewidth,height=.178\linewidth]{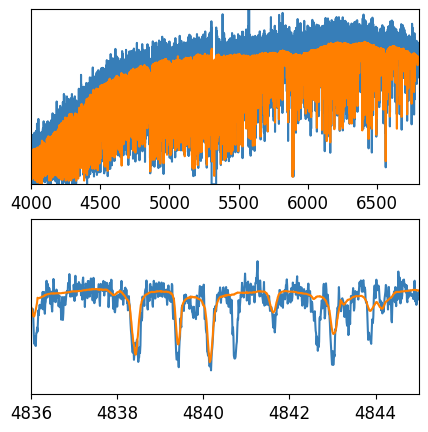}  
    & 
    \includegraphics[width=.2\linewidth,height=.178\linewidth]{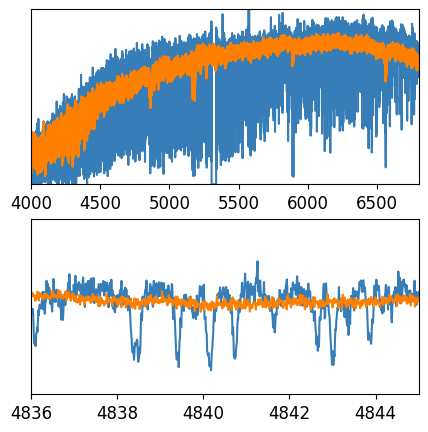}

    \\
    \smallskip
    \\
    \includegraphics[width=.2\linewidth,height=0.2\linewidth]{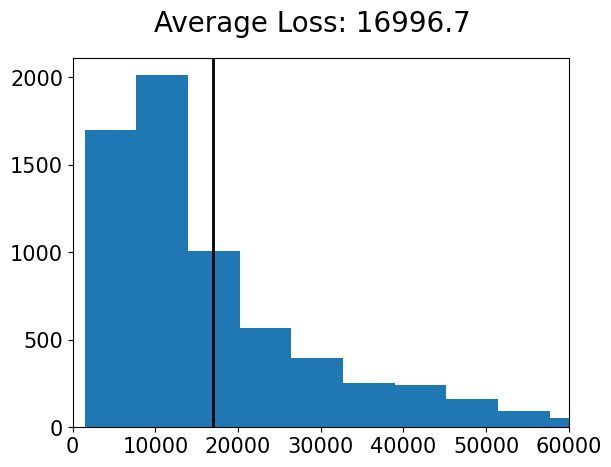}
    & 
    \includegraphics[width=.2\linewidth,height=0.2\linewidth]{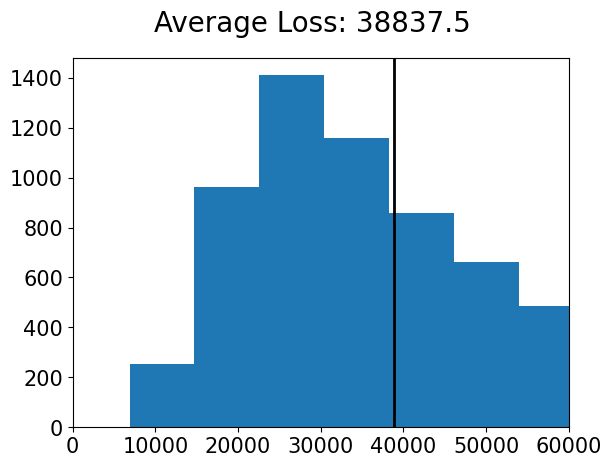}  
    &
    8-D
    &
    \includegraphics[width=.2\linewidth,height=0.178\linewidth]{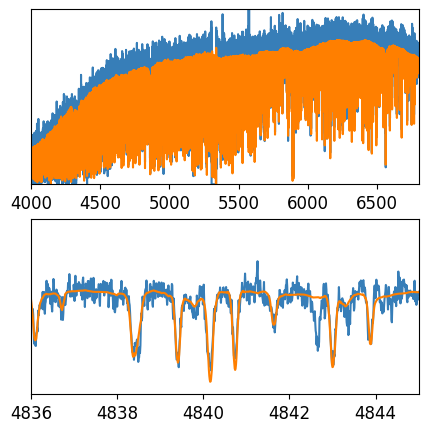}  
    & 
    \includegraphics[width=.2\linewidth,height=0.178\linewidth]{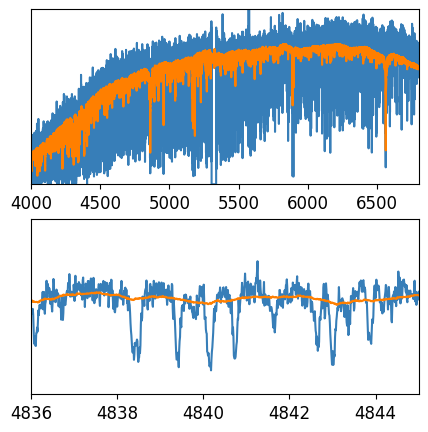}
    \\
    \smallskip
    \\
    \includegraphics[width=.2\linewidth,height=0.2\linewidth]{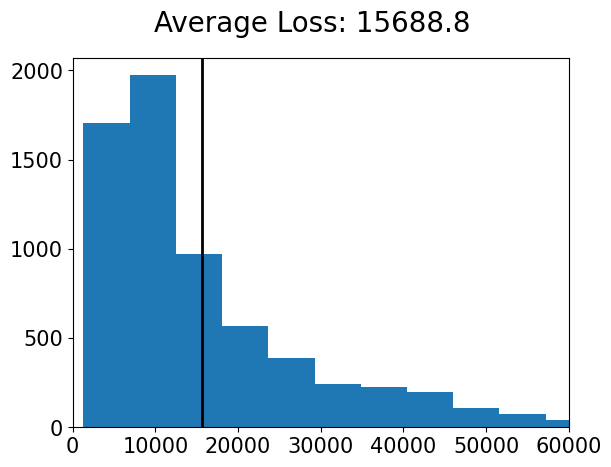}
    & 
\includegraphics[width=.2\linewidth,height=0.2\linewidth]{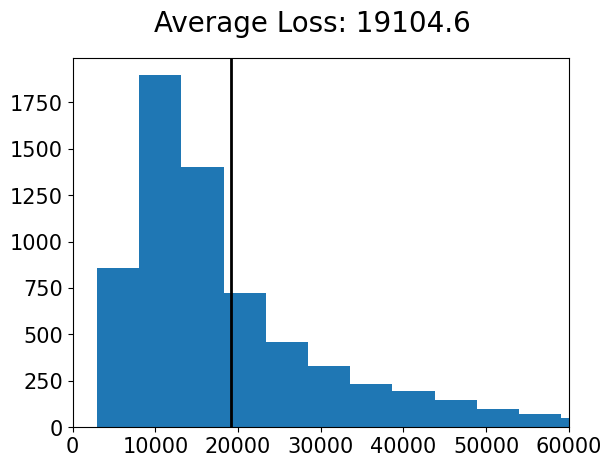}  
    &
    128-D
    &
    \includegraphics[width=.2\linewidth,height=0.178\linewidth]{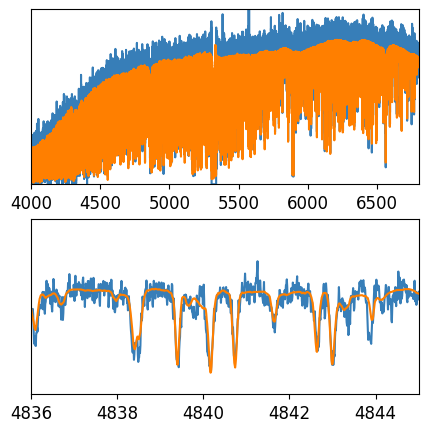}  
    & 
    \includegraphics[width=.2\linewidth,height=0.178\linewidth]{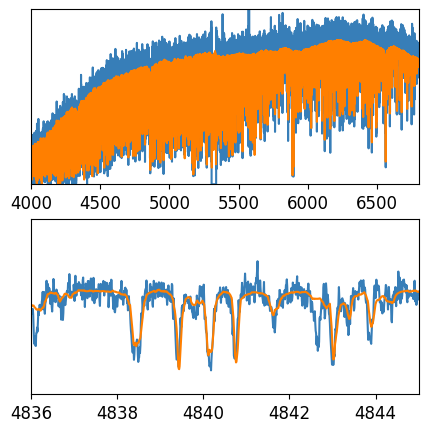}
    \\
\end{tabular}
  \caption{Illustration of the effects of two major factors on reconstruction quality: latent space dimensionality and disentanglement. 
  The left two columns illustrate reconstruction loss over the whole spectra, while on the right the same effects are depicted, in two different zoom levels, on an exemplar single spectrum: Input(blue) and reconstructed version(orange) are overplotted.
  Comparing the results of the deterministic autoencoder, and that of the disentangled variational autoencoder, we can clearly see the sacrifice in reconstruction quality, that occurs for the sake of disentanglement.
  On the other hand, as we increase the number of latent dimensions (top-down direction in the figure), reconstruction quality for fine details is enhanced.}
  \label{fig:reconst}
\end{figure*}


\section{Dataset}
\label{sec:dataset}
The dataset is built from observations using the HARPS instrument, a fibre-fed high-resolution echelle spectrograph dedicated to the discovery of exoplanets \citep{2003Msngr.114...20M}. 
The spectrograph has a resolving power of 115,000 and covers the spectral range 378--691nm.
We use the $\sim$270000 HARPS fully reduced spectra available in the ESO Science Archive\footnote{The retrieval form to access these spectra is at \url{http://archive.eso.org/wdb/wdb/adp/phase3_main/form}} in our investigations.

The datatset consists primarily of stellar spectra, although has an extended diversity due to the presence of solar system objects such as Jupiter and its Galilean moons, and asteroids. Although these objects are potential contaminants, we decide to leave them in the dataset, to keep the degree of supervision close to zero. We only had to remove unusable spectra: the ones containing undefined or unrealistic flux values, reflecting instrumental errors.

The spectra are homogenized by trimming down to the same minimum (3785 \AA) and maximum (6910 \AA) wavelengths, and then zero-padded either side to the reach the same number of pixels. We chose this length to be $327680 = 2^{18}+2^{16}$ -- reasonably close to a power of 2 for computational purposes. With the same resolution (0.01 \AA), the wavelengths in the spectra are therefore represented by the index of the flux vector. The result is a 1-dimensional input for the network to train on.

\subsection{Imbalanced Observations}
\label{balancing}
Any dataset can potentially have different numbers of observations (instances) for different objects. An extreme example in the case of HARPS is HD128621 (\textalpha{} Cen B) for which there are $\sim$20000 instances in the dataset, whereas many other objects have been observed only once.

Just like in any other data-driven method, ignoring this effect, which is quite similar to a \textit{selection function}, would allow dominant objects to inject bias and prevent the learned features from being representative of the whole dataset. But in order to stay fully unsupervised we take two parallel approaches and compare the results: First we implement a \textit{visibility balancing} technique in which visibility weights are incorporated during training, set to be inversely proportional to the occurrence frequency of each object in the dataset. Then we also run the same experiments ignoring the imbalance. 

As we will see in the upcoming sections, the major physical concepts that are captured by the network remain consistent across the two experiments. However, as expected, some other nodes start to learn features influenced by the dominant (class of) objects.

Also, in some of the test experiments, we are interested in looking at each object only once. We extract a `unique' list of objects for this purpose, in which multiple observations of each object are discarded and simply the first one is picked. We extract the number of occurrences only based on the `target-name' field in the database. While the target names in HARPS are not $100\%$ reliable, we decide to accept the error as it can only influence the results in a negative way, and does not introduce any kind of false hope. In the 272376 spectra queried from the database at the start of the work \footnote{We make the subset available to public.} we get 7653 unique target names.

\section{Reconstruction Results}
\label{sec:reconst}
\subsection{Deterministic AutoEncoder}
Theoretically, the quality of the reconstructed spectra should heavily depend on the size of the bottleneck, as it reflects the amount of preserved information.

Reconstructions with various bottleneck sizes are displayed in \cref{fig:reconst}. 
Interestingly, with a bottleneck as low as 8 dimensions, we already get a very good reconstruction of most of the spectra.

With only two latent dimensions, the network tends to preserve only the overall shape of the spectrum. Conversely,
the higher the number of bottleneck dimensions is, the more accurately the output follows fine features of the input. 
A detailed analysis of this behaviour is beyond the scope of this paper and will be provided in an upcoming article.

\subsection{With Disentangled Features}
\Cref{fig:reconst} also depicts reconstruction examples with disentangled features.
As expected, disentanglement comes at the cost of losing reconstruction quality. Hence, to obtain a high degree of reconstruction quality and disentanglement at the same time, the bottleneck needs to have a higher number of dimensions.

\subsection{Training Set vs. Validataion Set}
We split HARPS into training and validation subsets simply based on the index, after being sorted on the `ADP ID' field. The field presents just a unique identifier and does not have any meaningful correlation with real-world features, such as observation time or object type, and is therefore safe for the purpose.

The split has been used to monitor the training process and avoid overfitting. We also investigated possible differences in reconstruction quality across the two subsets subjectively, and found no meaningful difference.

\section{The Physics the Network Learns to Infer} 
\label{sec:physics}
The main objective is not for the network to reconstruct the input with a high accuracy, but rather to learn a minimal \textit{useful} representation of the spectra. 
In this section we try to interpret the learned features, and seek to find traces of physical semantics.
We pursue \textit{ablation study} by cracking open the network and analyzing the statistical behaviour of the latent nodes. 

To this end, we forward-pass an ensemble of spectra half-way through the network and store the ensemble of latent representations, to form a $n \times d$ matrix of \textit{codes}. This compact matrix, in practice, contains the whole ensemble, in a compressed format, and suffices for all statistical analyses. 
We use the unique subset introduced in \cref{balancing} for this purpose, since dominant objects in the dataset, like \textalpha{}Cen-B with \ntild{20000} instances, would bias and occlude our analyses otherwise.

\begin{figure}[]
    \centering
\begin{tabular}{ c c c }
    \tiny{$\lambda=0.03$} & \tiny{$\lambda=0.3$}
    & \tiny{$\lambda=0.9$}
    \smallskip
    \\
    \includegraphics[width=0.3\linewidth]{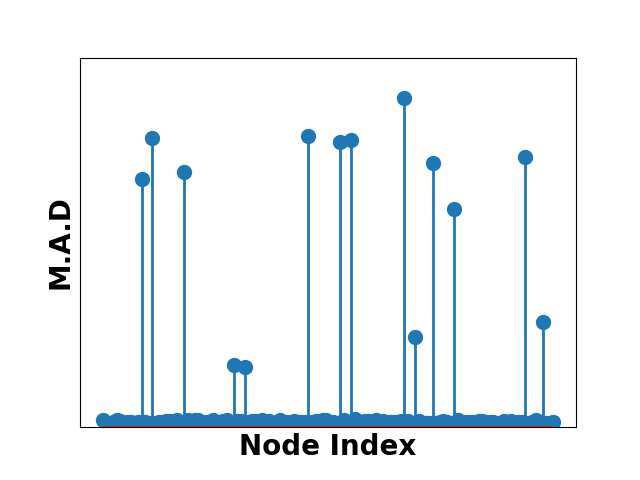}
    &
    \includegraphics[width=0.3\linewidth]{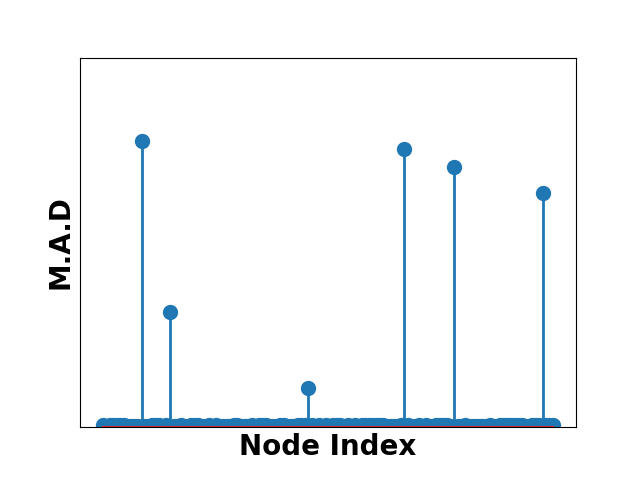}
    &
    \includegraphics[width=0.3\linewidth]{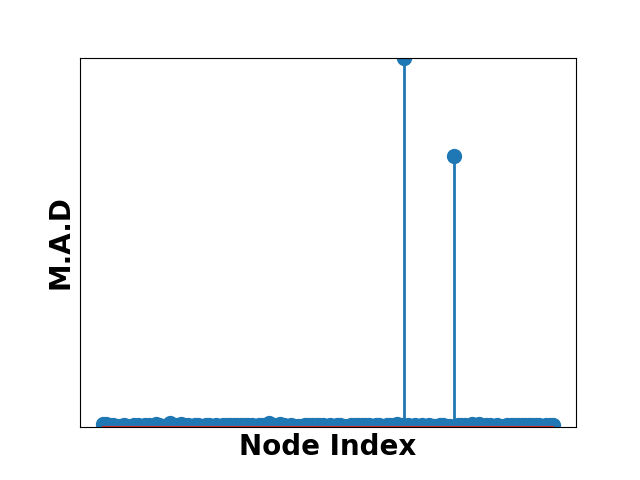}
    \\
    &
    \includegraphics[width=0.3\linewidth]{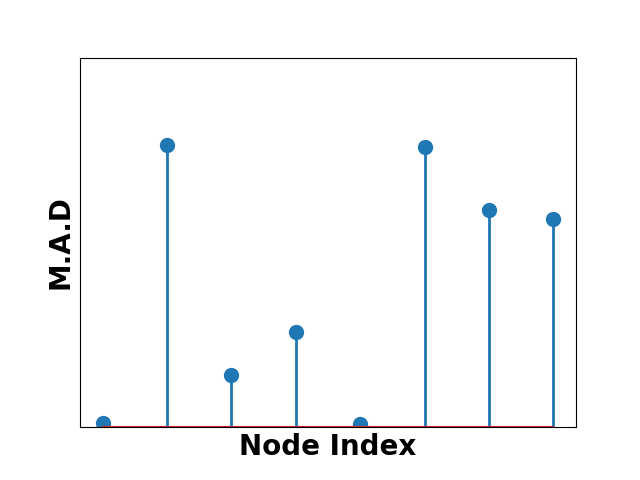}
    &
    \includegraphics[width=0.3\linewidth]{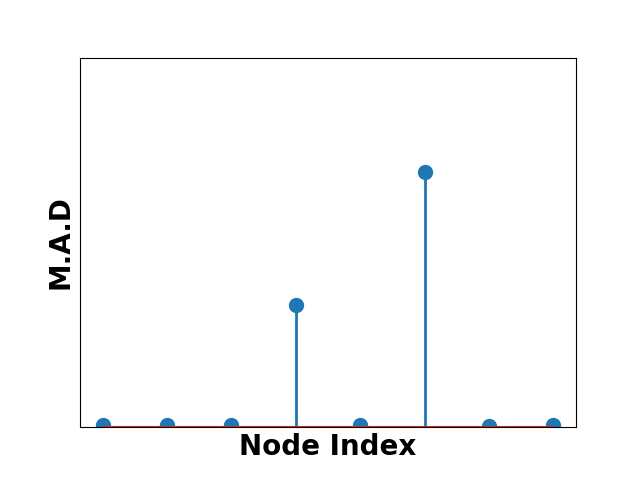}
    \end{tabular}

    \caption{M.A.D values for 128-d network on the top and 8-d network on the bottom. From left to right the disentanglement weight ($\lambda$) is increased. Too low weights result in leak of information among different dimensions, while too high values cause loss of details which causes better disentanglement, yet less useful features. Interestingly the 128-d and 8-d networks agree on the number of informative features at $\lambda=0.3$.}
    \label{fig:mads}
\end{figure}

\begin{figure}[]
    \centering
    \begin{subfigure}{0.6\linewidth}
        \centering
        \includegraphics[width=\linewidth]{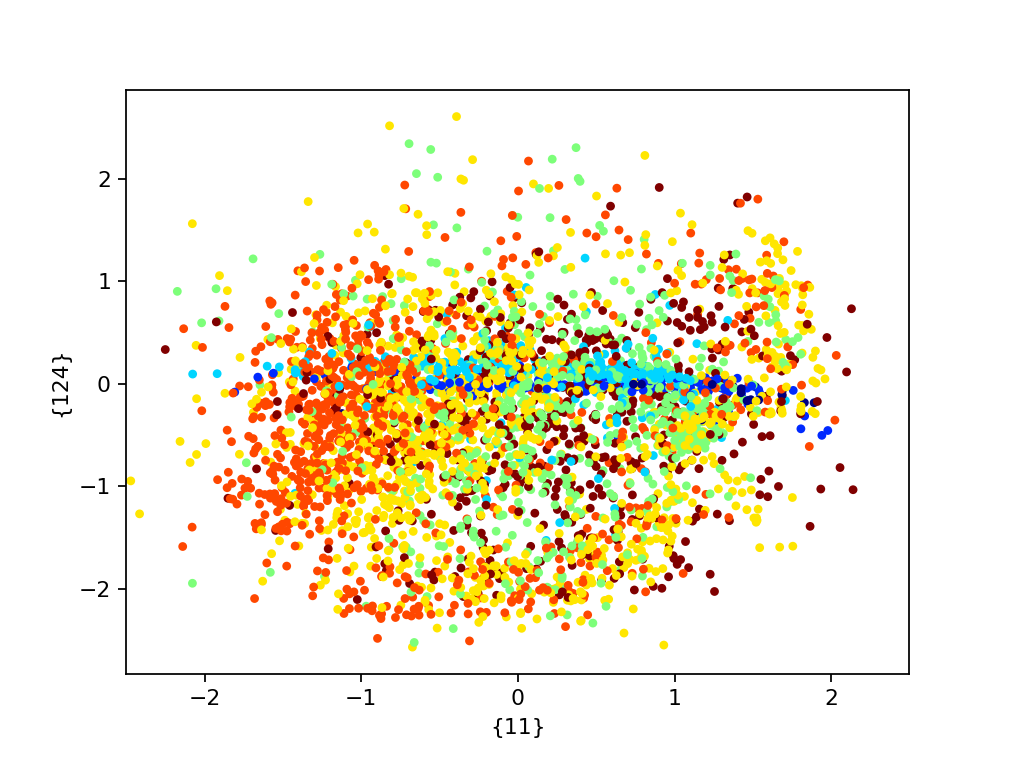}
        \caption{$\lambda=0.3$}
        \label{fig:xcorr_success}
    \end{subfigure}
    \begin{subfigure}{0.6\linewidth}
        \centering
        \includegraphics[width=\linewidth]{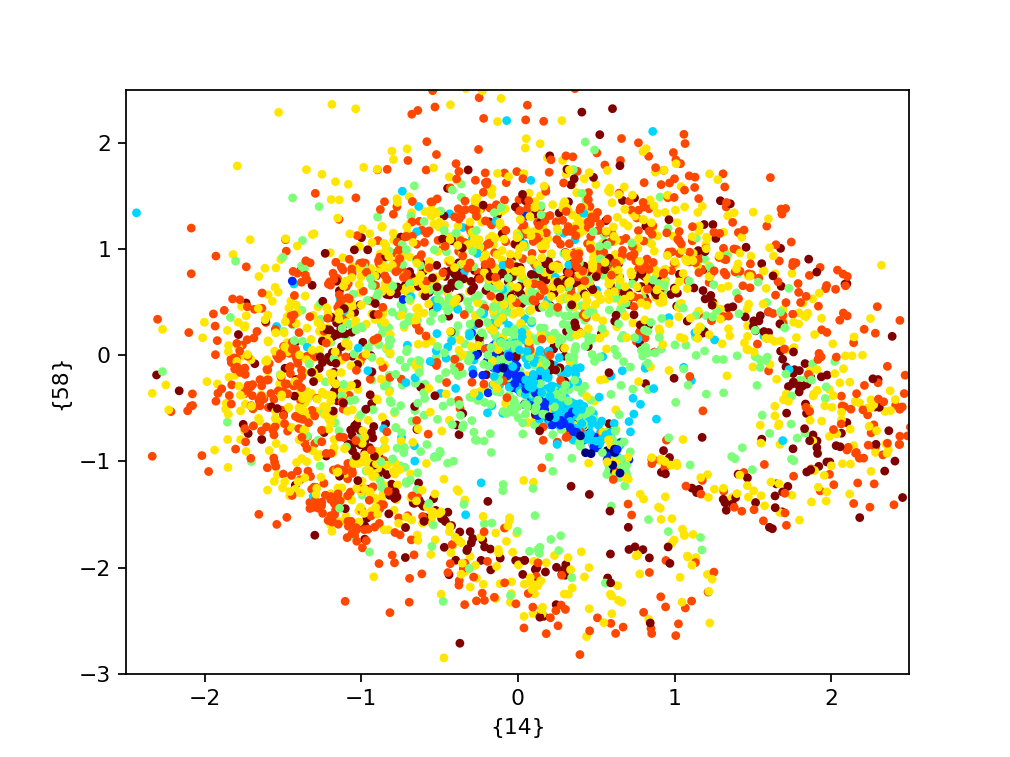}
        \caption{$\lambda=0.1$}
        \label{fig:xcorr_entangled}
    \end{subfigure}
    \begin{subfigure}{0.6\linewidth}
        \centering
        \includegraphics[width=\linewidth]{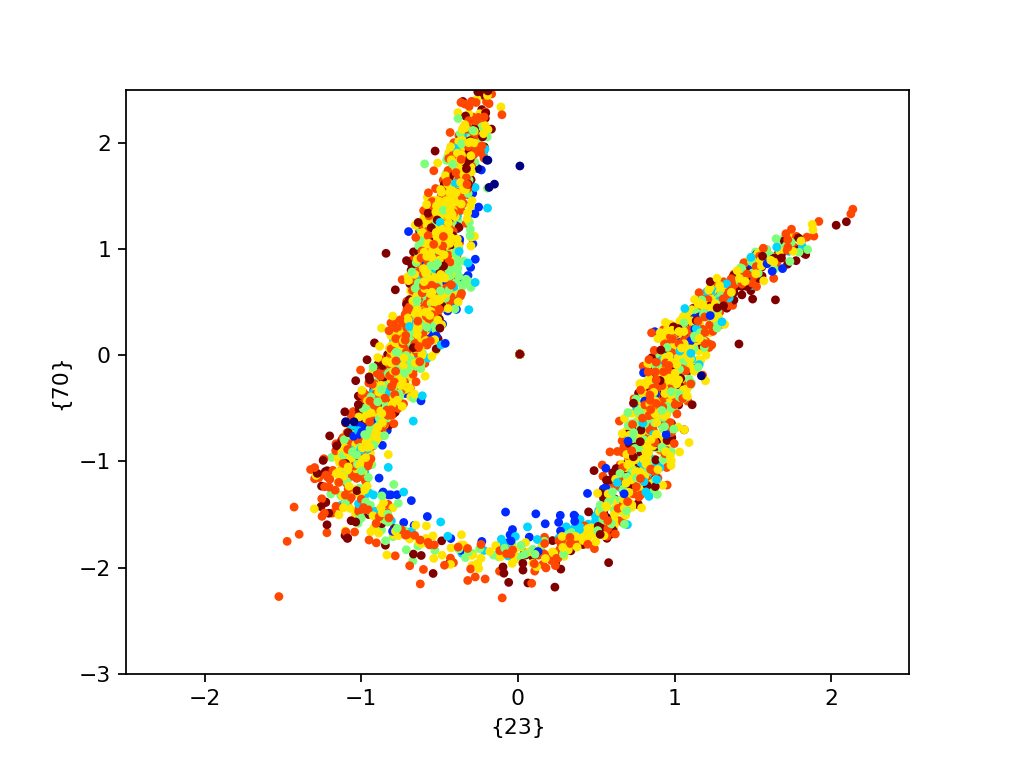}
        \caption{$\lambda=0.03$}
        \label{fig:xcorr_entangled2}
    \end{subfigure}
    \includegraphics[width=0.6\linewidth,height=18pt]{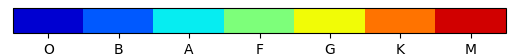}

    \caption{Scatter plots illustrating mutual behaviour of pairs of latent dimensions. On the top, there is little to no significant correlation between the two. In contrast, the bottom two plots show clear correlation between exemplar dimension pairs, in networks where $\lambda$ has been too low, which is a strong hint for failure of disentanglement. In such cases, a high M.A.D does not directly translate to possession of exclusive information. Contrary to intuition, the less structured the plots are, the more successful the disentanglement has been.
    Different colors show different spectral classes and are used for illustration purposes only.}
    \label{fig:xcorr_example}
\end{figure}

\begin{figure}[]
\parbox[c][1.05\linewidth]{0.02\linewidth}{
\centering
\vspace{-10pt}

\rotatebox{90}{\tiny{$\lambda=0.3$}}
\\ \vfill

\rotatebox{90}{\tiny{$\lambda=0.5$}}
\\ \vfill

\rotatebox{90}{\tiny{$\lambda=0.9$}}
}
\parbox{0.8\linewidth}{
  \begin{center}
    \includegraphics[width=\linewidth,height=0.5\linewidth]{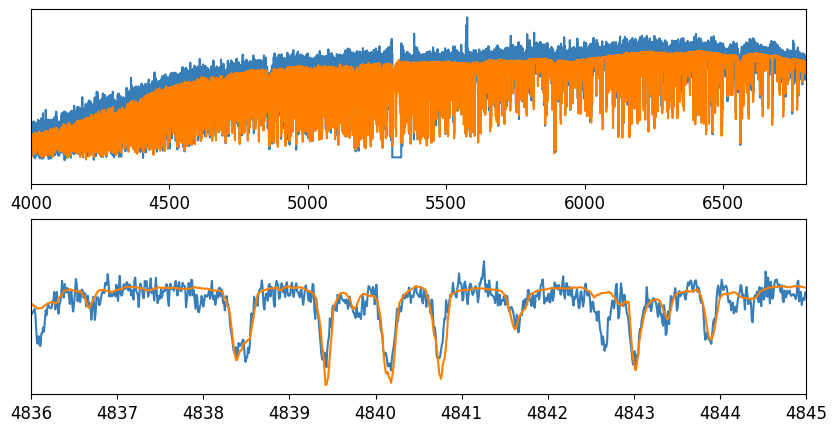}
    \\\bigskip \bigskip
    \includegraphics[width=\linewidth,height=0.5\linewidth]{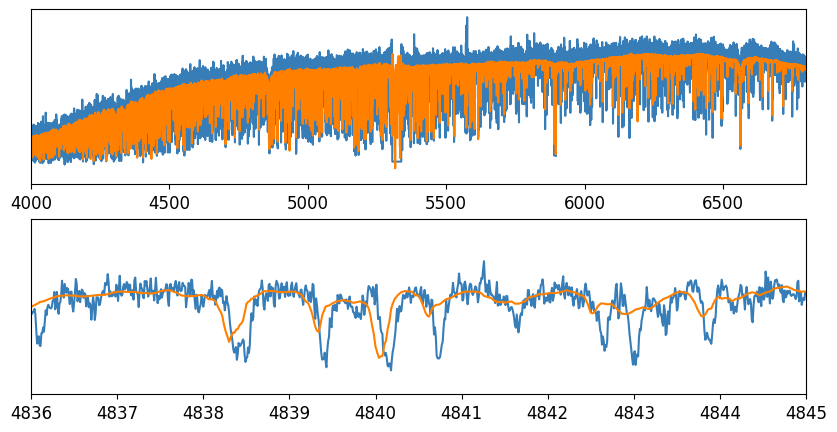}
    \\\bigskip \bigskip
    \includegraphics[width=\linewidth,height=0.5\linewidth]{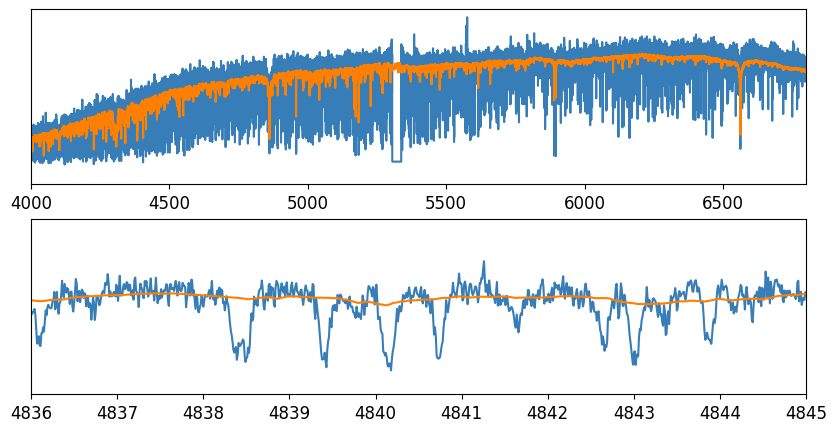}
    \tiny{wavelength (\AA)}
  \end{center}
  
}
    \caption{From top to bottom, the effect of too much disentanglement enforcement is visualized. The network loses the ability to preserve details, i.e. narrow lines, and starts focusing on the overall shape only. In such a case, although the significant dimensions learn disentangled representations, the captured concepts are too simplistic and not much useful.}
    \label{fig:too_much}

\end{figure}

\subsection{Informative Dimensions}
Our very first analysis is to find out how many \textit{informative} features the network really has learned. To this end we utilize median absolute deviation ($MAD$), as a robust measure of statistical dispersion as an initial score of informativeness. The score for the $i^{th}$ latent node ($Z^i$) is computed as: 

\begin{equation}
    \operatorname{MAD}^i =\underset{j}{\operatorname{median}} (|Z^{i}_{j}-{\tilde {Z^i}}|)
\end{equation}

\noindent where $j$ iterates over samples (spectra) and  ${\tilde {Z^i}}=\underset{j}{\operatorname{median}}(Z^i)$.

Although such a dispersion measure is, by definition, tied to the diversity of the underlying dataset, still any \textit{important} property of the samples should show enough variability across different samples -- or else it contains close to zero \textit{information} for our purpose, hence deemed unimportant.

There is one degree of freedom (hyper parameter) which seems to affect the number of informative nodes: the disentanglement weight ($\lambda$) of \cref{eq:beta-vae-loss}. In \cref{fig:mads} we see that, lower levels of disentanglement simply result in too many \emph{significant} dimensions, which cannot be called informative anymore, as disentanglement is not really happening. \Cref{fig:xcorr_example} depicts how two significant dimensions may still be highly correlated -- evidence that the disentanglement has failed.

Too much disentanglement, on the other hand, results in fewer significant dimensions, which may seem as a good outcome in the first look. However, our experiments show that reconstruction quality decays so much that fine details are discarded and the few learned features are all centered around the overall shape of the spectra -- \cref{fig:too_much}. This trade-off is a well-studied characteristic of unsupervised disentanglement methods -- \citealp[e.g. see][]{burgess_understanding_2018}. Networks with other bottle-neck dimensionalities follow the same trend, although narrower bottlenecks inherently tend to (have to) discard fine details.

We find that a disentanglement weight $\lambda$ of around $0.3$ provides a reasonable trade-off, where no two significant dimensions show significant correlation -- i.e. good disentanglement. Interestingly, we find exactly 6 informative latent dimensions in two different networks with latent dimensionalities of 8 and 128. 

\begin{figure*}[t]
  \begin{center}
    \parbox{\linewidth}{

    \parbox{.9\linewidth}{%
    \begin{flushright}
    \hfill\rotatebox{0}{\tiny{Rad. Velocity}} \includegraphics[trim=178 500 140 500,clip,width=0.9\linewidth]{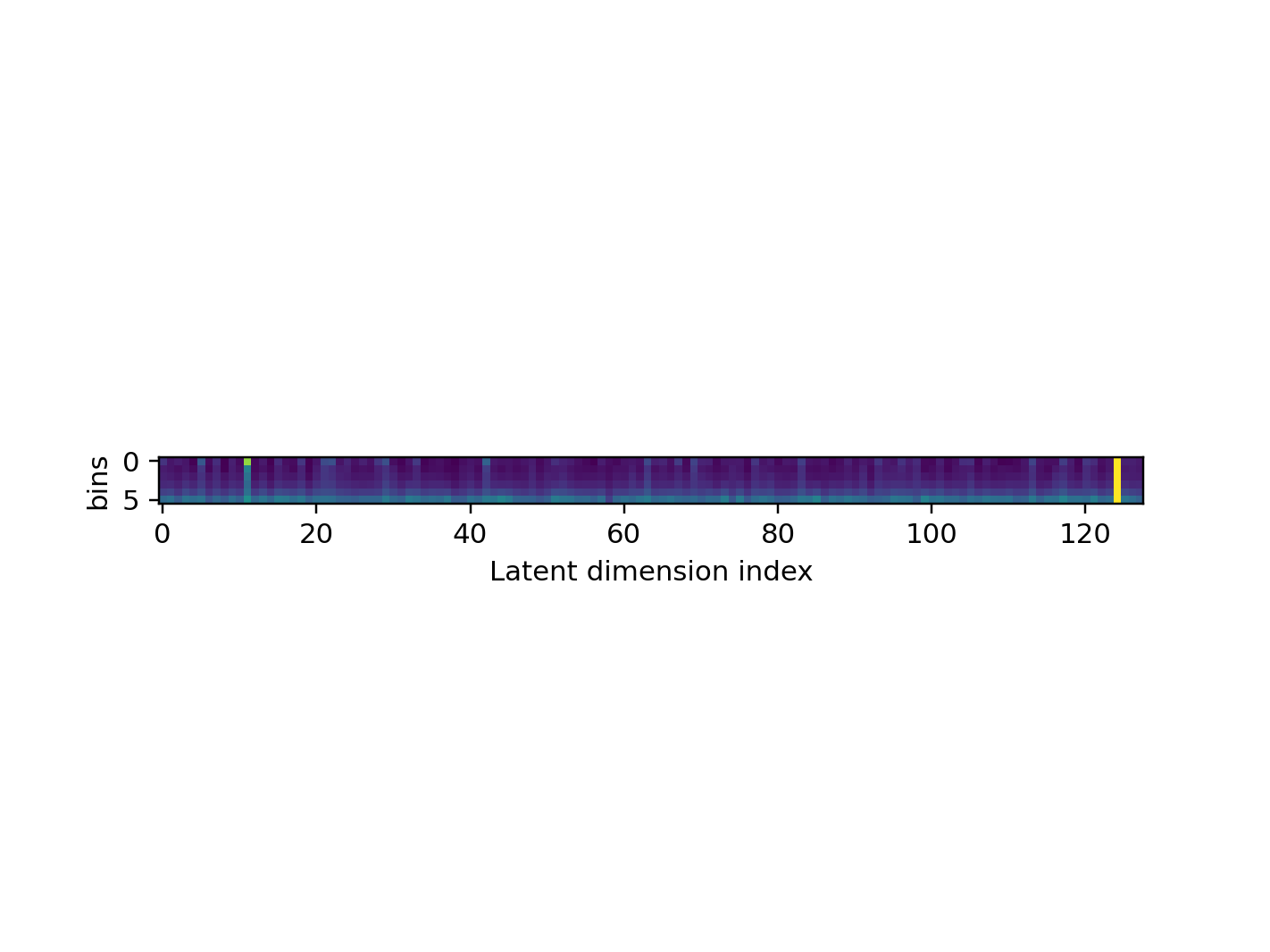}\\
    \hfill\rotatebox{0}{\tiny{Eff. Temp.}} \includegraphics[trim=178 500 140 500,clip,width=0.9\linewidth]{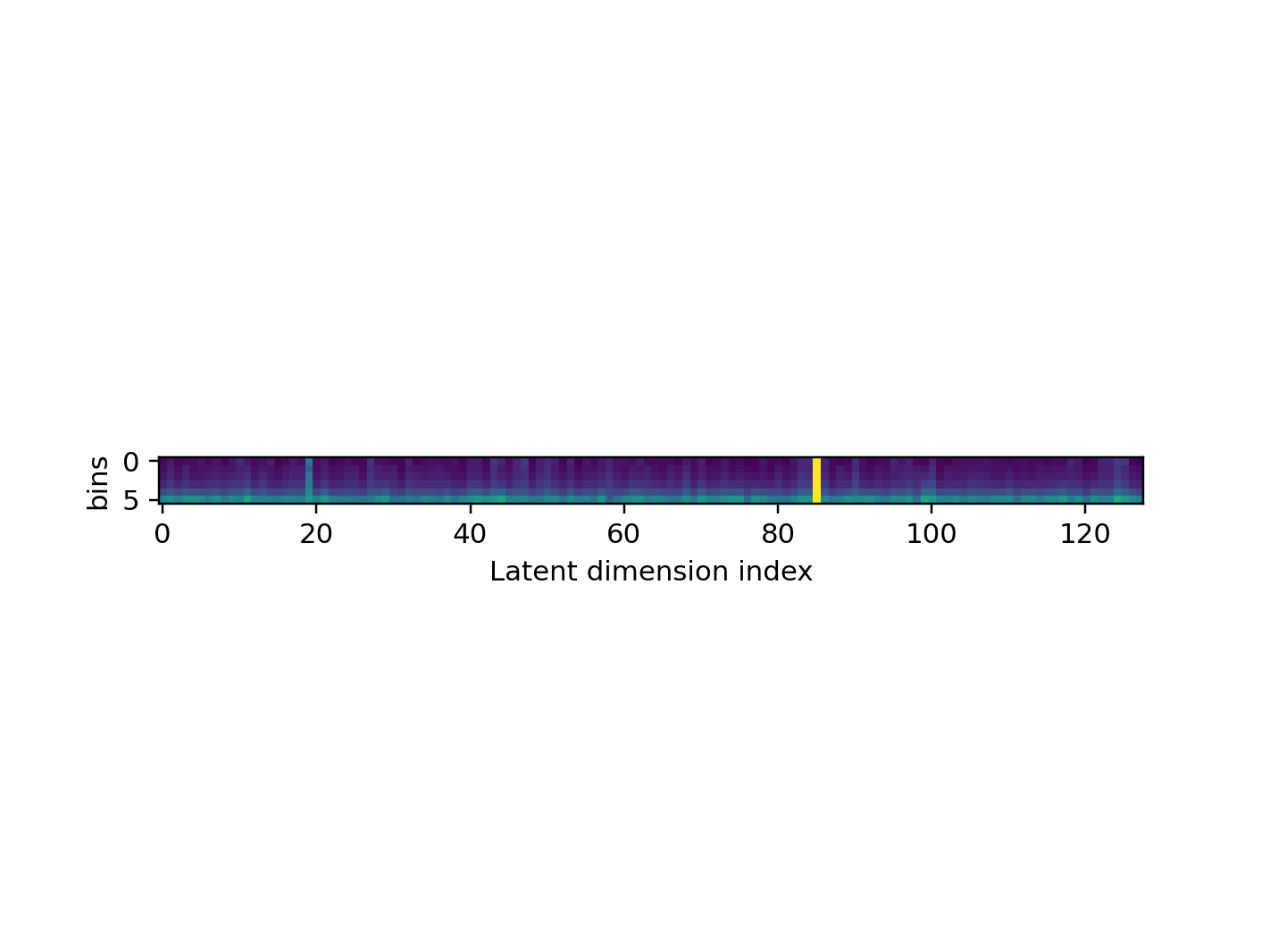}\\
    \hfill\rotatebox{0}{\tiny{Surf. Gravity}} \includegraphics[trim=178 500 140 500,clip,width=0.9\linewidth]{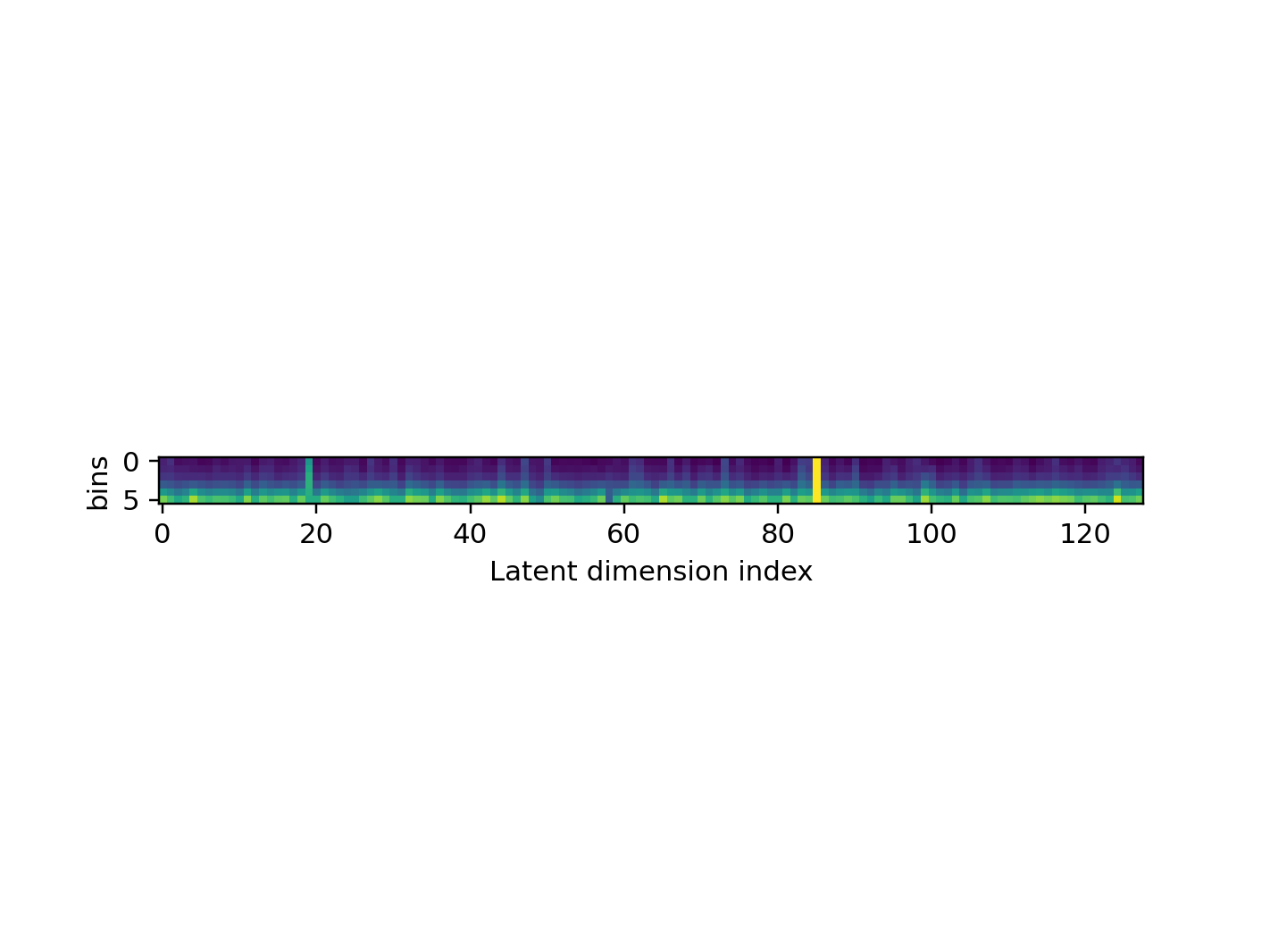}\\
    \hfill\rotatebox{0}{\tiny{Metallicity}} \includegraphics[trim=178 500 140 500,clip,width=0.9\linewidth]{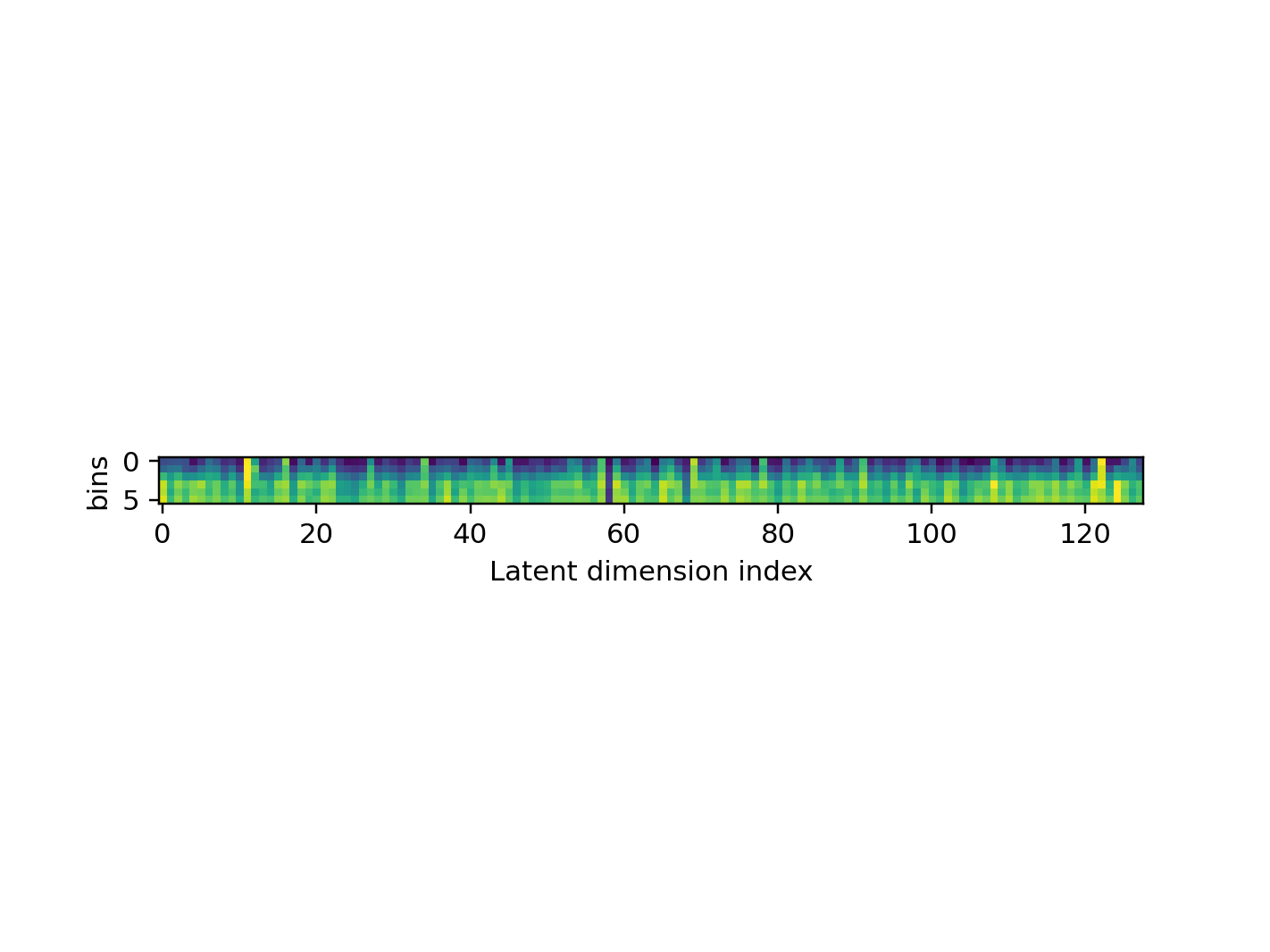}\\
    \hfill\rotatebox{0}{\tiny{Airmass}} \includegraphics[trim=178 500 140 500,clip,width=0.9\linewidth]{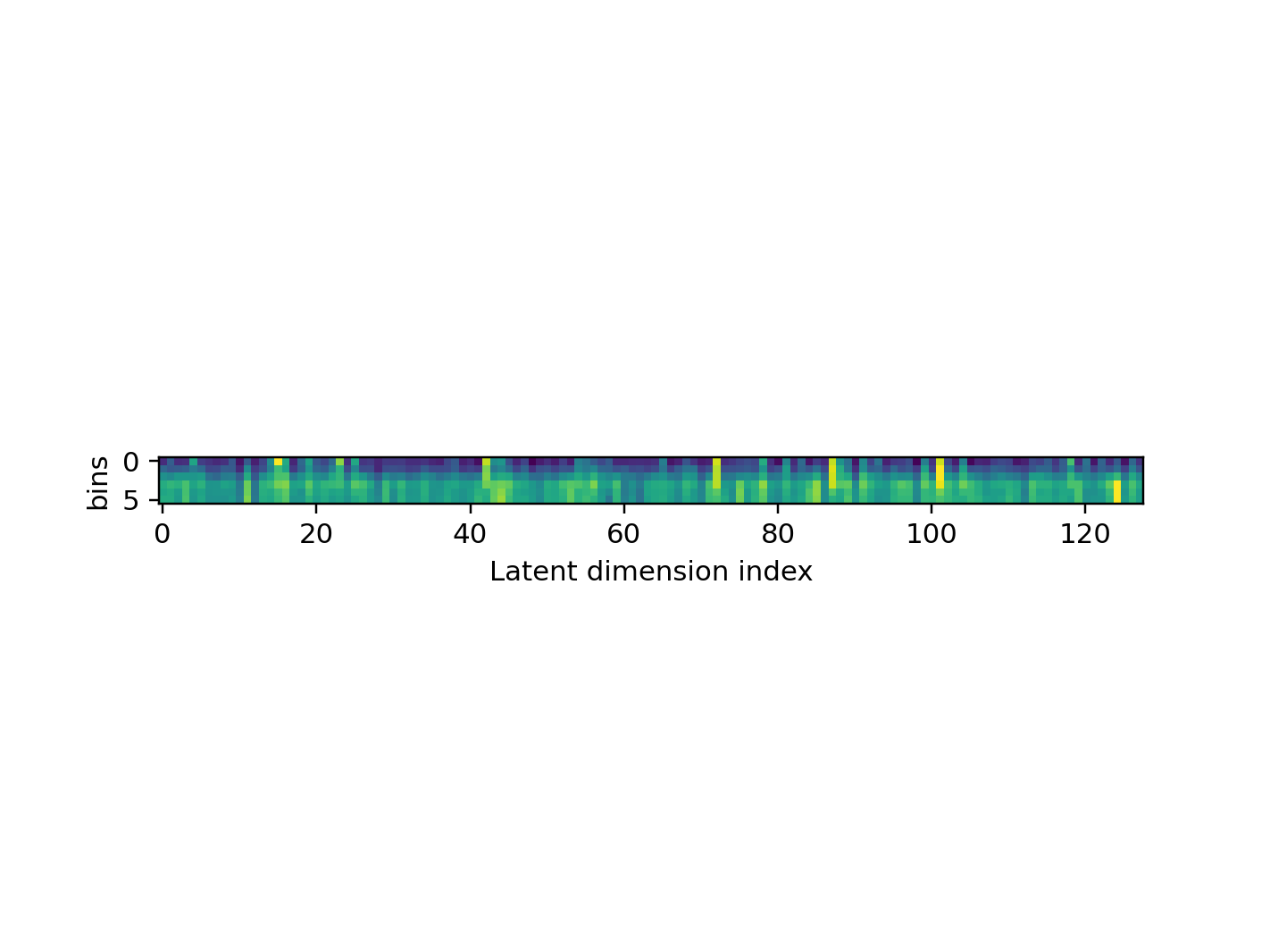}\\
    \hfill\rotatebox{0}{\tiny{SNR}} \includegraphics[trim=178 500 140 500,clip,width=0.9\linewidth]{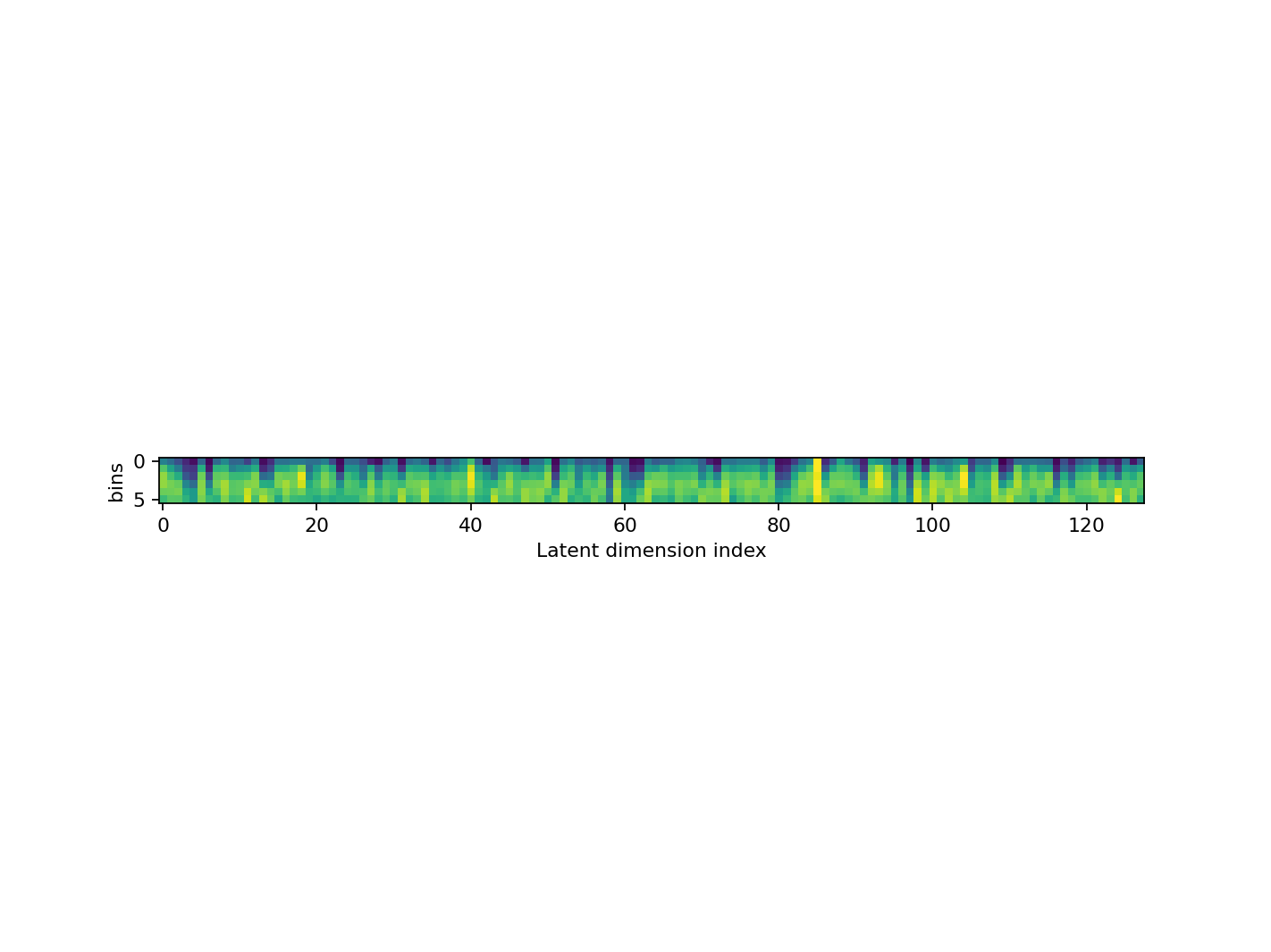}
    
    \end{flushright}
    }
    \parbox{.04\linewidth}{%
      {\includegraphics[width=0.59\linewidth]{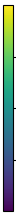}}
    }
    }  
    \tiny{$\longleftarrow$ Latent Node Indices $\longrightarrow$}
  \end{center}
  \caption{Correlation indicators based on Mutual Information at different scales. The depicted matrix at each row shows different scales (binning configurations) along the vertical axis and different nodes are sitting horizontally. Each row of each indicator, representing a single scale, is normalized by max.
  For Radial Velocity, Effective Temperature and Surface Gravity, individual nodes stand out, while for Metallicity, Airmass and SNR, that is not the case.}
  \label{fig:MIs}
\end{figure*}

In the next section, we take an information theoretic approach towards detection of traces of physics in latent features, which is completely independent of the the informativeness indicator of this section. But as we move forward we find a reassuring harmony between the two methods.

\subsection{Mutual Information -- with Known Physics}

So far we have identified the dimensions which, from a purely statistical point of view, seem to have captured \textit{significant} features of the stars. Now we seek to interpret the learned features and find specific traces of physics. The search is conducted over all the latent features, to avoid any bias from the statistical scores of previous section.

Assuming we have access to a large number of known \mbox{(astro-)physical} parameters, we seek \emph{Mutual Information} between them and the latent features the network has learned.
Pearson correlation is too limited as it can only capture linear dependence with Gaussian noise, while "Mutual Information is able to quantify the strength of dependencies without regard to the specific functional form of those dependencies" \citep{kinney_equitability_2014}.

Mutual information of two jointly discrete random variables is defined as \citep{cover_elements_1991}:
\begin{equation}
    I(X;Y)=\sum_{y\in{\mathcal{Y}}}\sum_{x\in{\mathcal{X}}}{p_{(X,Y)}(x,y)\log {\left({\frac {p_{(X,Y)}(x,y)}{p_{X}(x)\,p_{Y}(y)}}\right)}}
\end{equation}

\noindent A more intuitive formulation is given by
\begin{equation}
\label{eq:mi_entropy}
         I(X;Y)=H(X)-H(X|Y)=H(Y)-H(Y|X)
\end{equation}

\noindent and defines Mutual Information as the amount of uncertainty lost in one of the variables by knowing the other one. In \cref{eq:mi_entropy}, $H(.)$ is the \textit{Shannon Entropy} \citep{shannon_mathematical_2001}.

\begin{figure}[t]
  \begin{center}
    \includegraphics[width=\linewidth]{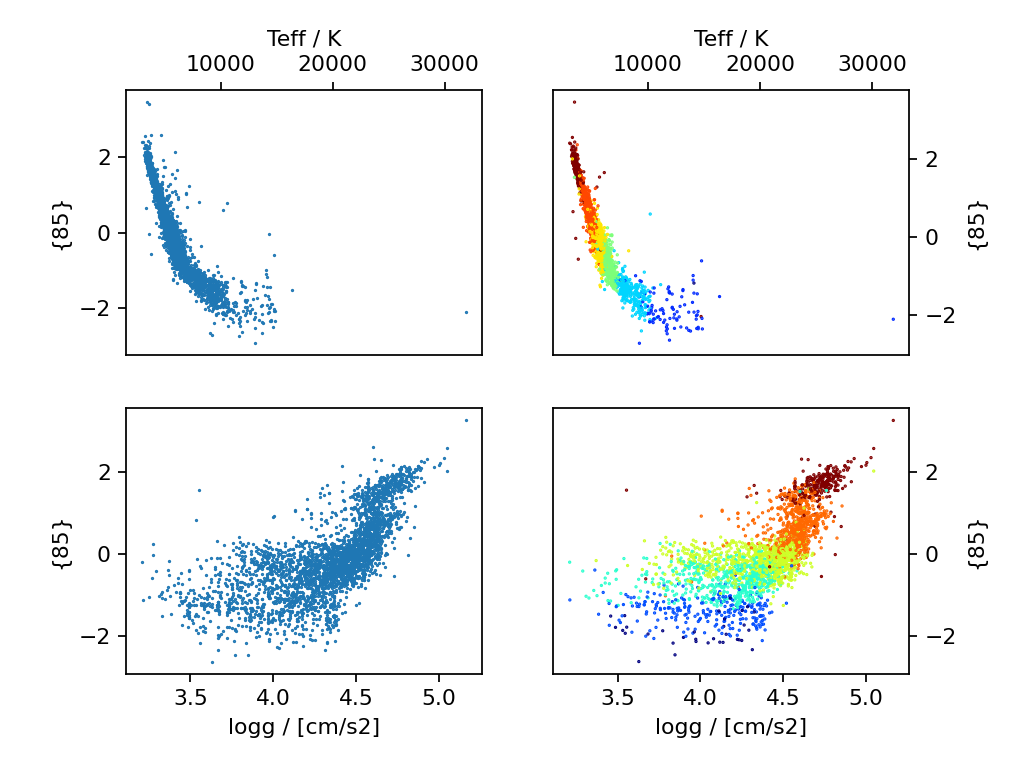}
    \includegraphics[width=0.5\linewidth,height=15pt]{material/colourbar.png}
    \includegraphics[width=0.7\linewidth]{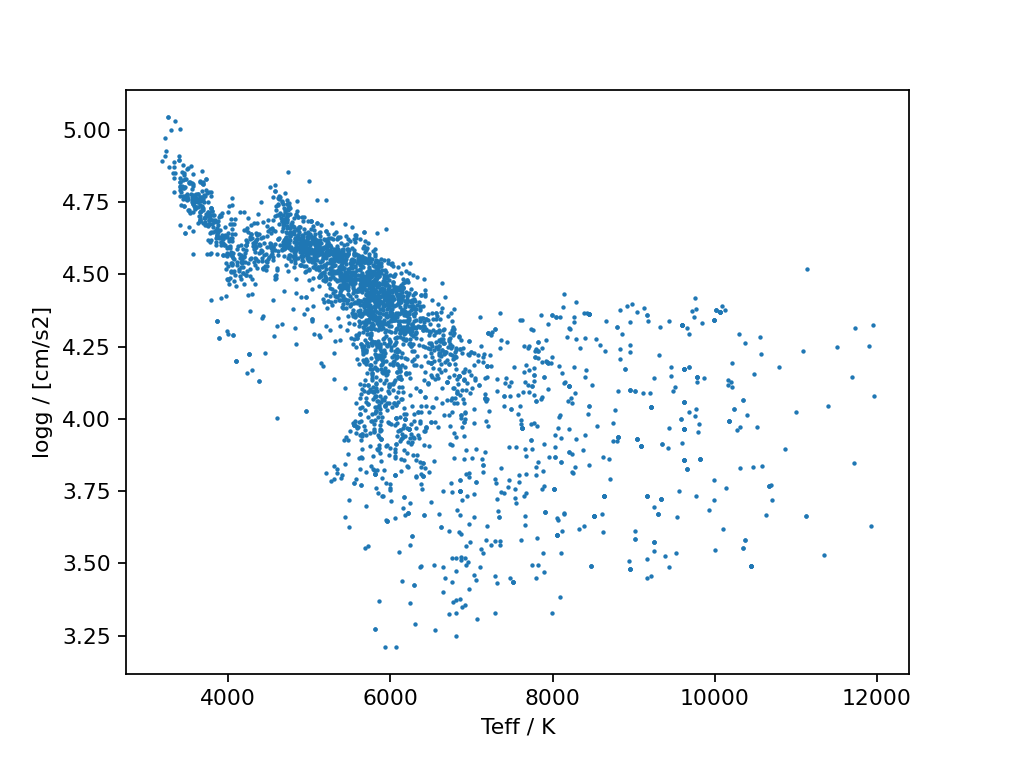}
  \end{center}
  \caption{Node \{85\} shows a good correlation with effective temperature -- top row. The tightness of the structure reflects the strength of the mutual information. The same node shows a not-so-strong correlation with surface gravity -- middle row. Plotting log(g) vs. Teff in the bottom row reveals the reason. Please refer to the main text for a detailed analysis. It is also useful to note that our sample is very biased towards main sequence stars, with the $\log g$ only varying between $\sim 3.5$ to 5.}
  \label{fig:feature_85}
\end{figure}

\begin{figure}[t]
  \begin{center}
    \includegraphics[width=\linewidth]{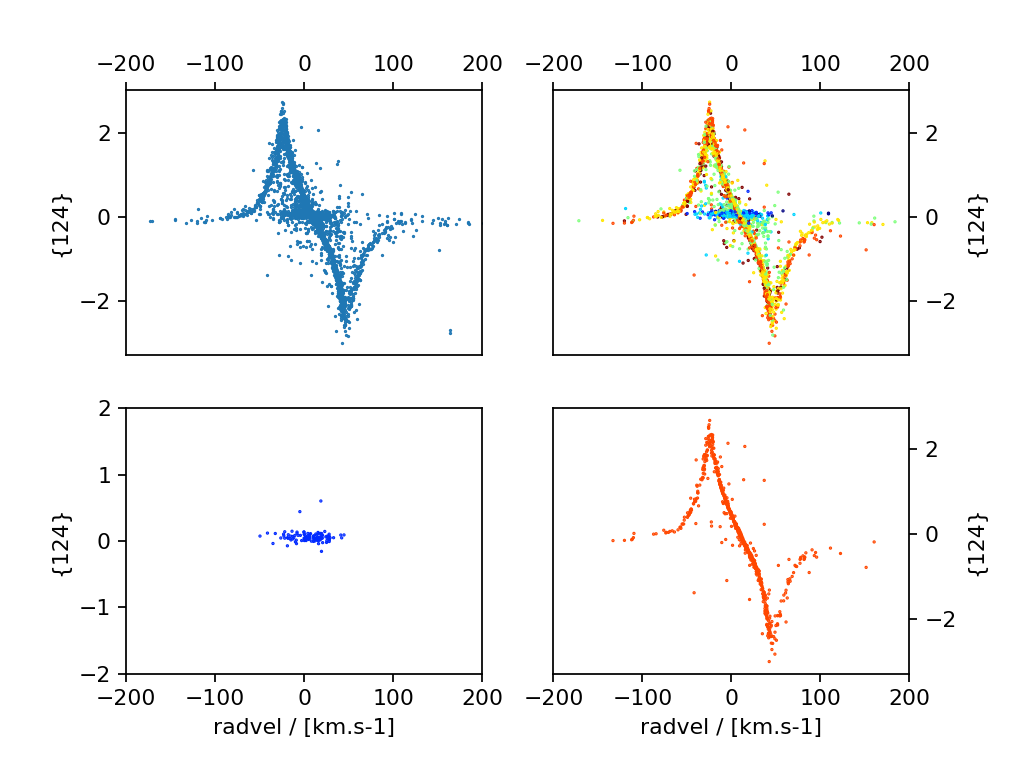}
  \end{center}
  \caption{Node \{124\} learns a clear understanding of a notion of radial velocity -- top row. The symmetric shape, and the fact that the network has automatically gained an understanding of \textit{zero velocity} as a reference point are notable observations. Different temperatures have apparently been treated differently, as also detailed in the bottom row.}
  \label{fig:feature_124}
\end{figure}

Given a number of data points, it is often difficult to obtain an accurate estimate of the Mutual Information (MI) of the underlying random variables, as it involves estimation of the underlying joint distribution. For the task at hand, however, we are not much interested in the exact value of the MI, as it is a relative indicator when considering all latent dimensions. 

We use joint histograms to simply approximate the joint density. Still, the estimated MI's turn out to be quite sensitive to the chosen number of bins. Therefore, to have a simple, yet robust indicator, we provide a 2-step workaround:
a) sigma-clipping at $5\sigma$, and b) multi-scale (scan at various bin resolutions).

We extracted some of the known astrophysical features, for a portion of our dataset, from SIMBAD \citep{wenger2000simbad}, TIC \cite{stassun2019tic}, and observation-time parameters:

\begin{itemize}
    \item Effective Temperature (Teff)
    \item Surface Gravity (log(g))
    \item Metallicity ([M/H])
    \item Radial Velocity
    \item Airmass
    \item Signal-to-Noise Ratio (SNR)
\end{itemize}

\noindent Steps of the process are detailed in \cref{app:Retrieve}\footnote{We re-emphasize that the learning process has been a fully unsupervised one and such labels have been merely used post-training for validation purposes only.}.

We construct MI indicators as explained above, to seek traces of these intrinsic astrophysical stellar parameters in all dimensions of our networks. Results for the 128-dimensional network are illustrated in \cref{fig:MIs}.
Clear signs of strong correlation are seen for Radial Velocity at dimension \{124\}, and Teff, log(g) at dimension \{85\}. No clear dimension stands out for [M/H] airmass and signal-to-noise ratio (SNR).

The two detected "physical dimensions" have already been identified by the purely statistical indicator of the previous section, which increases the reliability of the finding.
Visualization of the direct relationship between latent features and their corresponding validation labels in \Cref{fig:feature_85,fig:feature_124}, shows that the network has clearly grasped a direct notion of these physical concepts.

\subsubsection{Analysis}
Node \{85\} shows correlation with both effective temperature and surface gravity. Its correlation with the effective temperature is clear, monotonic and tight, providing close to a one-to-one mapping from node values to temperatures -- \Cref{fig:feature_85}, top row. 

The reason surface gravity is captured with the same dimension, becomes clearer after plotting the scatter of the two physical parameters (not the node values) against each other -- bottom row of \Cref{fig:feature_85}. It turns out that the input dataset presents a biased view when it comes to temperature and gravity, in that it does not sample uniformly the general underlying stellar population. Concretely speaking, in the objects the network has seen, temperature and surface gravity are more or less strongly correlated. From an information theoretic point of view, surface gravity does not provide much exclusive information, and a big fraction of the information in it is shared with effective temperature. In other words, the network does not need to dedicate an independent node to store information about this physical parameter, when it can obtain most of what it needs from another node -- especially under disentanglement pressure. Of course, the network needs to store the exclusive part of the information about this parameter, which is reflected in the scattered points in the plot, somewhere. That place is most likely in one of the discarded nodes.

Node \{124\} has captured information on the stars' radial velocity. The correlation is shown visually in \cref{fig:feature_124}. The plot shows that the network has automatically learned a model for hypothetical, reference, zero-velocity spectra, since it has formed a symmetric mapping around it. The mapping is of course not a bijective function. It is also worth noting that for colder stars the correlation is quite tight and progressively loosens for hotter stars, until it essentially vanishes at the highest temperature available in our dataset. We speculate that the increasing sparseness of absorption features with increasing temperature is responsible for the observed behaviour.

The spectral absorption from the Earth atmosphere as parametrized by the airmass affects the large-scale shape of the spectra, a prominent feature that could be expected to be picked out by the network. The same could be expected for metallicity.
A posteriori, however, this does not seem to be the case since neither of these parameters are significantly correlated with any of the dimensions, as gauged by the MI results, which may look puzzling at first glance.
This may be, however, related to the fact that HARPS has a relatively narrow wavelength range, mostly bluewards of most telluric features. HARPS is mostly an exoplanet hunter, and those are mostly looked at around solar-like or cooler stars, and our sample is strongly biased against containing early-type stars. This can be seen in Fig.~\ref{fig:feature_85}, where it is also clear that our dataset is mostly comprising main-sequence stars. It also covers a limited range in metallicity, while the optimised {\tt New Short Term Scheduler} used by most HARPS visitors implies that most targets are observed at the best (i.e. lowest) airmass possible. It is therefore not surprising that the algorithm could not find a correlation with metallicity and airmass.

One may also expect SNR to be captured by the network as an independent feature, since it plays a role in forming the appearance of an spectrum. This is, however, not the case and comes as little surprise; the noise is uncorrelated with any other type of information in the dataset and by definition does not contain any \textit{pattern} across different spectra to be learned. Thus, for a model to capture and reconstruct pixel-accurate noise, it would need to assign one parameter per pixel per spectrum -- i.e. memorize the noise. This advantageous limitation is a well-known feature of even the simplest classical autoencoders, such that denoising autoencoders have been among the first ones to be used \citep{vincent2010stacked}. Such behaviour is of course seen in many other methods used for dimensionality reduction, such as PCA -- e.g. see \citet{bailer1998automated}.

\subsection{Latent Space Traversal}
Although we run out of available physical labels or/and automatically detected correlations, we go further and pursue deeper investigation based on a method known as \textit{Latent Space Traversal}. 
\begin{figure*}[]
  \begin{center}
    \raisebox{-0.5\height}{\includegraphics[width=0.2\linewidth]{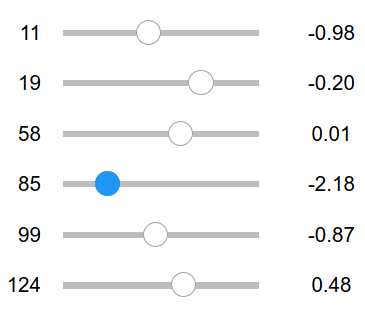}}
    \raisebox{-0.5\height}{\includegraphics[width=0.75\linewidth]{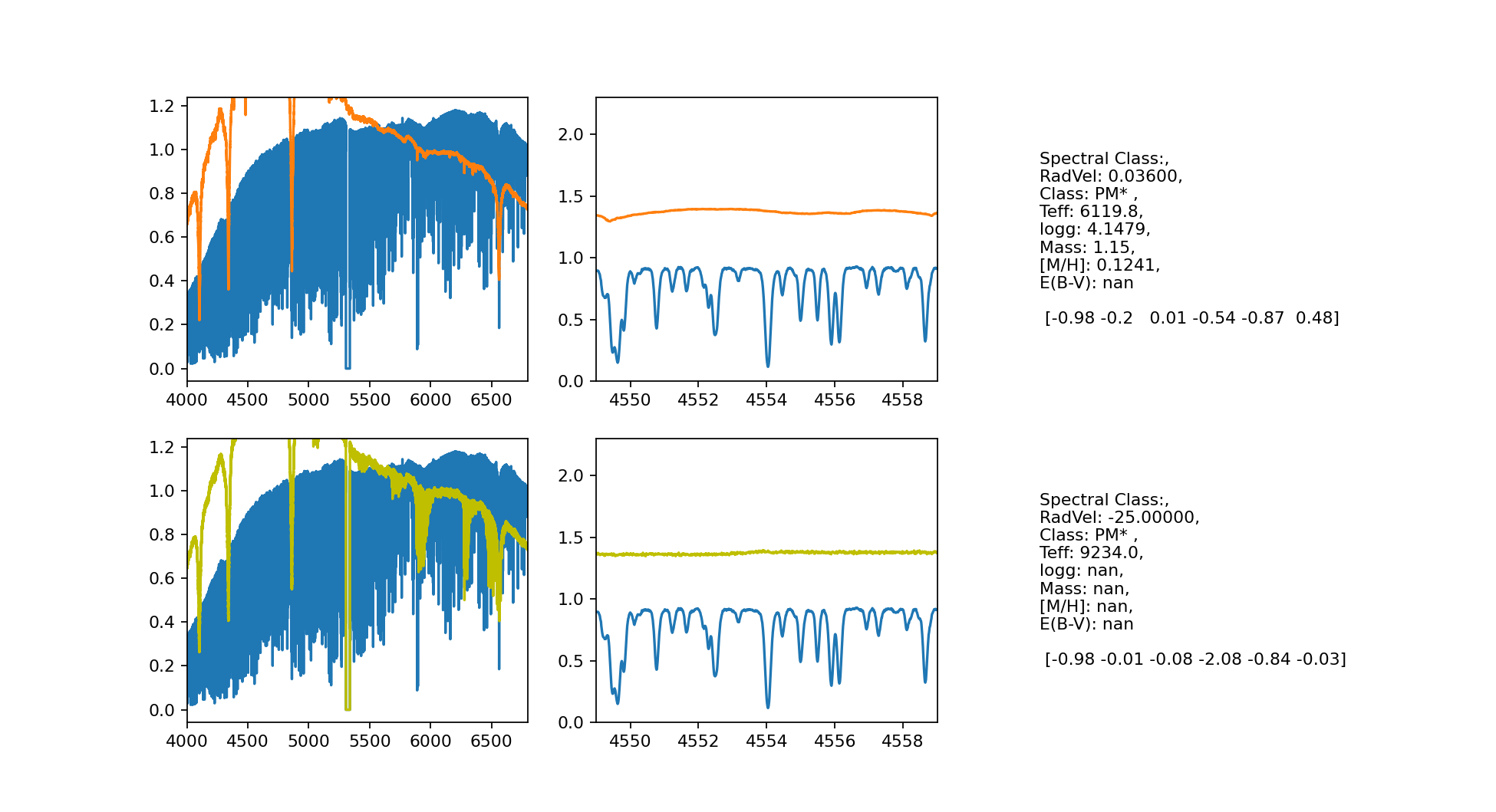}}
    \\
    \raisebox{-0.5\height}{\includegraphics[width=0.2\linewidth]{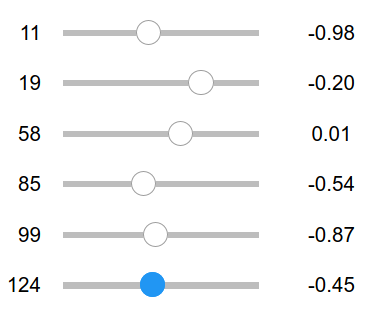}}
    \raisebox{-0.5\height}{\includegraphics[width=0.75\linewidth]{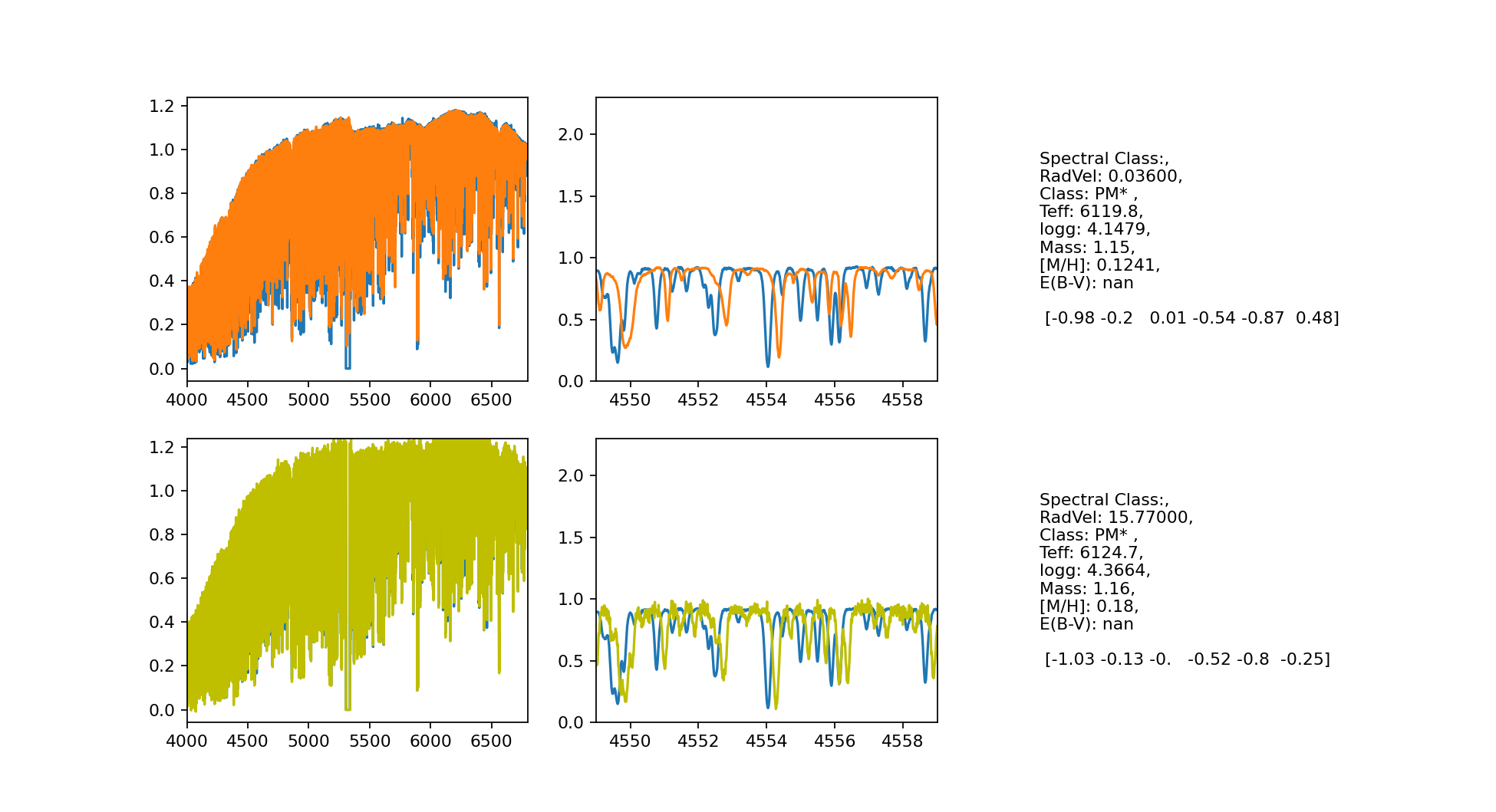}}    

    \bigskip 
    \includegraphics[width=0.45\linewidth]{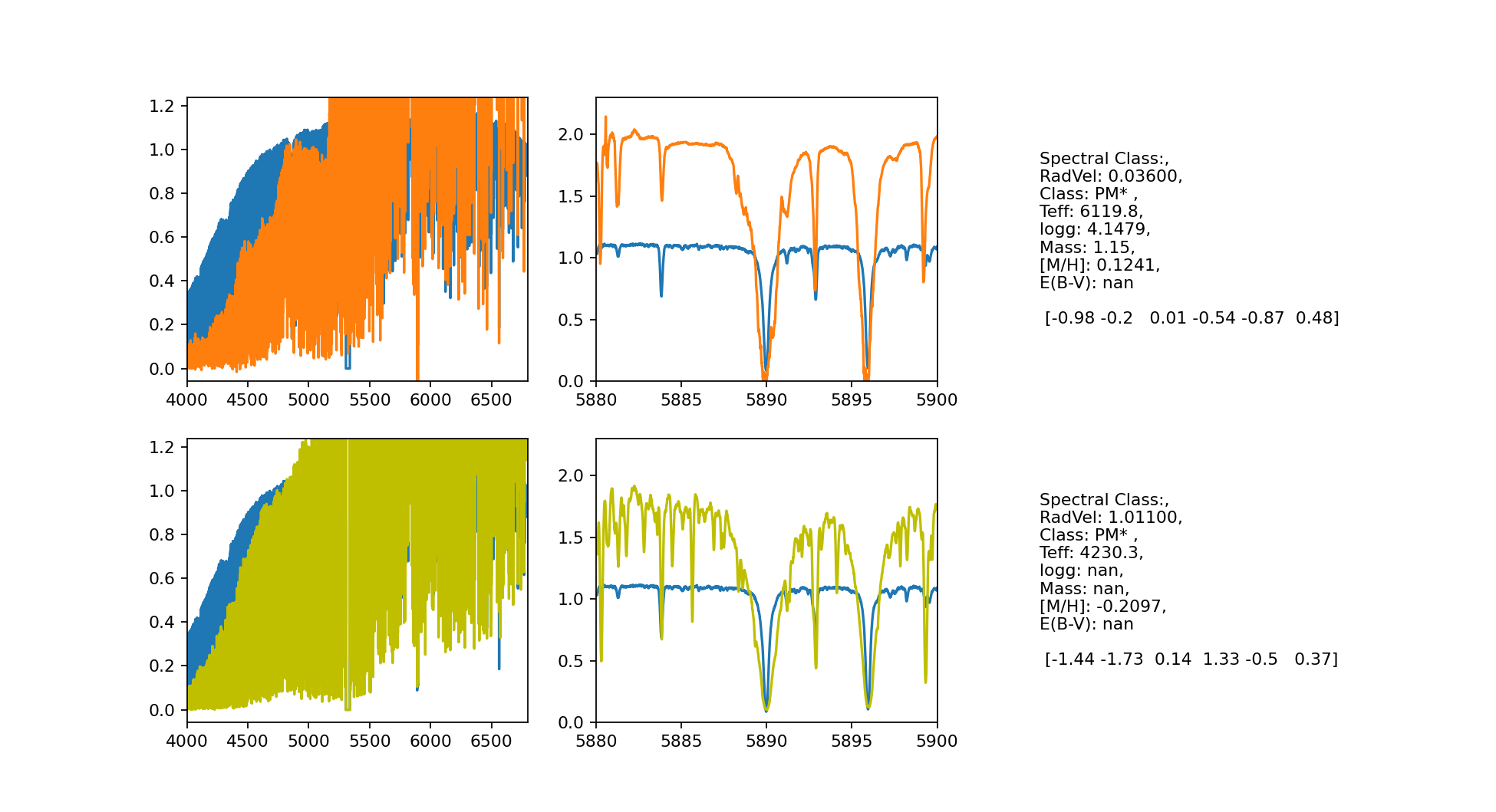}    
    \hskip0pt
    \includegraphics[width=0.45\linewidth]{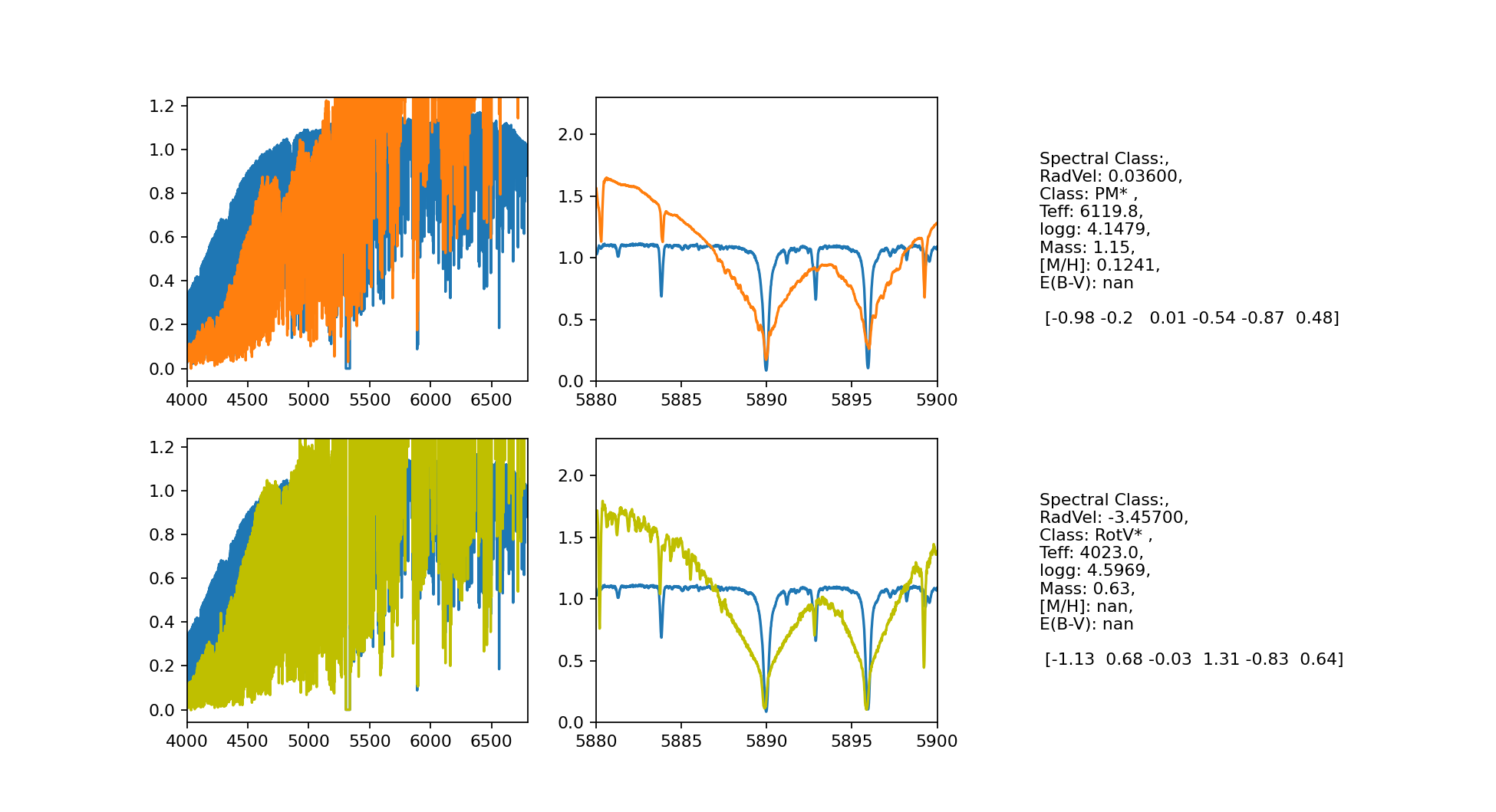}    
    \\
    \includegraphics[width=0.15\linewidth]{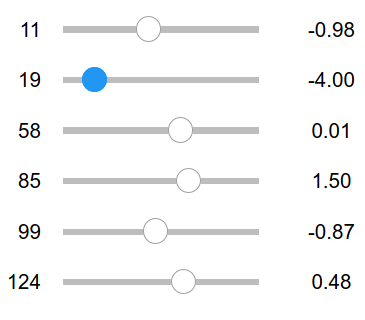}
    \hskip140pt
    \includegraphics[width=0.15\linewidth]{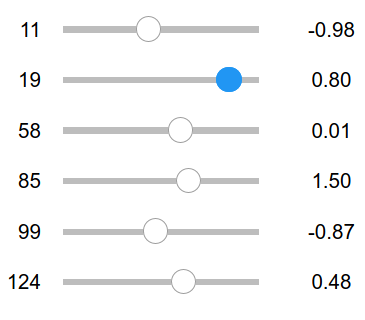}
  \end{center}
    \vspace{-10pt} 
    \caption{Our interface for latent space traversal, showing 3 different experiments. All experiments share the same randomly chosen `reference' star, shown in blue. This reference is encoded by the network, the obtained code is slightly modified using the sliders, and decoded to generate the orange spectrum. This resulting spectrum is usually an imaginary one and thus, we illustrate the closest real object to it in green. This closest object is searched for in the learned latent space. From top to bottom we show experiments for the effects of \embrace{85} (effective temperature), \embrace{124} (radial velocity) and \embrace{19} respectively. For the latter, which applies only to cool stars, we had to `move' the base spectrum to a late-type star, using \embrace{85}}
    \label{fig:traversal}

\end{figure*}
We start by forward-passing single spectra half-way through the network, just the way we did in the beginning of this section, to encode the spectrum into its latent representation. Then by perturbing (or traversing, in extreme cases) the code and generating the corresponding spectrum, we can have synthetic spectra which are different to the (reconstructed version of) the original spectrum as a result of the change in the code. So, singling out dimensions of the latent space allows for analysis of the effects of specific dimensions on the generated spectra, hopefully equal to interpretable features.

To this end we create an interface with sliders which allow for traversal over different dimensions and visualization of the effects on the fly -- \cref{fig:traversal}. In the following we list the significant findings. 

\paragraph*{Node \embrace{11}} seems to be, partly, related to the rotation of the star, which is another parameter that is known to affect the spectra -- a higher rotation will broaden the lines, making them less deep. Varying the value of this node does not affect the shape of the continuum, but only the depth of the lines.
Thus, an increased value of the node corresponds to much broader lines and this is clearly an effect of increasing rotational velocities (or macroturbulence in general). 
Above a given threshold, however, the situation is more complex:  for solar-like stars, the match seems to be done only on stars that have quite a large radial velocity shift. We have not yet found a physical reason for this. For early-type stars, the lines do not become broader either, but instead the Balmer lines clearly become narrower. This is likely an effect of the gravity of the star.

\paragraph*{Node \embrace{19}} is only affecting a subset of our sample, namely only the coolest stars. It has indeed no effect on solar-like stars or early-type stars, but only affects stars that have a value of node \embrace{85} above about 0.85, that is, stars cooler than $\sim$4500 K. For these stars, this node is clearly linked with the luminosity of the star. This node is thus also physical, but confined to a subset of the stars, only the coolest ones -- see \cref{fig:traversal} for an illustration.

\paragraph*{Node \embrace{58}} has, similarly to above, no apparent effect on the spectra of solar-like or early-type stars, but only manifests itself for even cooler stars than node \embrace{19}, those that are characterised with a value of node \embrace{85} above about 1.2.
However, we could not find as yet a clear explanation of the effect at play when varying the value of node node \embrace{58}, and we defer a detailed analysis to further work.

\paragraph*{Node \embrace{99}} is contrarily to node \embrace{11} affecting the continuum of the star, more than the lines themselves. It is also, unlike the previous two nodes, not really affecting solar-like and cooler stars, but has only a visible effect on stars hotter than the Sun.
From a phenomenological point of view, this node appears to be looking at the inflexion point of the continuum and whether the spectrum is thereby concave or convex. 
Thus, for very negative values (e.g. -4.5), there is a depression in the spectrum around 5800 \AA, which disappears at about $-1.7$, while for positive values, there is a maximum around 5300 \AA.
The clear physical explanation of this apparent phenomenological node is hard to find, but a first investigation indicates that it may be related to the presence of a disc (such as around Be stars) or a companion. Further studies are needed.

\subsection{Discarding Observation Frequencies}
Using the non-balanced dataset, we obtain 5 significant nodes, 2 of which correspond exactly to the major captured physical features: \embrace{85} and \embrace{124}.

Two nodes represent  features that are also seen as in the balanced set: \embrace{11} and \embrace{88} (the latter, corresponding to node \embrace{99} of the balanced set).

Representations captured in nodes \embrace{19} and \embrace{58} of the balanced net are clearly not present any more, as we cannot spot any node specifically representing only the coolest stars. 
The effect of the remaining node, \embrace{99}, is not clear cut. It seems that for the hottest stars ($> 10,000$ K), it is partially sensitive to the gravity of the stars: the lowest values of this node correspond to white dwarfs (i.e. high gravity), while the highest values correspond to solar-like stars. It has no apparent effect for A/F/G stars, nor for M stars, but there is an effect on K stars as well. We could not identify the physical nor phenomenological criteria that would correlate with this node.  

\section{Conclusion}
We implemented the idea of ``letting the data speak for itself'' in action, in the context of an astrophysical application, where we let a deep convolutional neural network look at stellar spectra and learn from them without any predefined objectives in mind. We showed that the network "chose to" learn how to extract and capture specific physical parameters of stars, among other unidentified ones, as their canonical features. The importance of the finding is in network's answer to ``what is important to learn?'', and should not be confused with the relatively trivial problem of training a network for estimation of those parameters.

Specifically, our purely statistical measure revealed that 6 out of 128 latent nodes of our network stand out as informative ones.
We also developed an information-theoretic indicator to track true/non-linear correlations between the learned features and a set of known astrophysical parameters. We found that two latent nodes, which interestingly turned out to be among the 6 informative ones, have clearly learned a notion of radial-velocity and effective temperature.

The automatic method did not indicate correlations between the remaining significant dimensions and the validation labels we had at hand. This does not necessarily indicate a false alarm on those nodes. They may have captured known physical parameters for which we do not have labels yet, or the existing labels might have not been quite reliable to reveal weaker correlations. 

Also, it is quite possible that the other nodes have not captured direct representations of familiar physical parameters, but rather other complex (or even simpler) features. Artificial neural networks do not have to think like humans! For example, We spot nodes which capture variations of specific absorption lines. They may have captured fine features of chemical abundances -- something that is not formulated in classical astronomy, with this level of granularity.
We believe such features that are not directly interpretable are interesting for follow up studies, since understanding the reasons behind a network's \textit{decision} to prioritize more complex/simpler features, or higher level relationships between basic features, may help advance our physical understanding of the underlying target -- stars in this case.

We continued with latent space traversal and found traces of rotation, luminosity, presence of a disc or a companion, in the unidentified nodes, some affecting only a subset of our sample (either the coolest stars, or the hottest). The latter correlations were, however, not as clear as the previous ones and were decided to be left for future studies. We make the interface available to public for this purpose.

As mentioned earlier, our dataset for this case study is very specific, due to the particularities of HARPS usage. It is to be expected that in more generic samples, other features, e.g., luminosity or metallicity, may come out more easily. In general, the concepts the network learns to capture, are dependent on the biases in the dataset.

\section*{Acknowledgements}
      This work is in part supported by the ESCAPE project (the European Science Cluster of Astronomy \& Particle Physics ESFRI Research Infrastructures) that has received funding from the European Union’s Horizon 2020 research and innovation program under the Grant Agreement n. 824064. We also acknowledge support for our research by funding from the Science and Technology Facilities Council.
      Lastly, we thank Michael F. Sterzik and Mark Allen for their help in preparation of the manuscript.

\section*{Data Availability}
    We release the code for the convolutional neural network, the list of IDs of the spectra used for training and validation and the physical validation labels on \url{https://github.com/NimSed/astro-machines}.
    We also make the "sliders" interface freely accessible to the community to facilitate study and discovery of new relationships with the introduced framework.

{\small
\bibliographystyle{mnras}
\bibliography{main}
}


\begin{appendix}
\begin{figure*}[th!]
  \begin{center}
    \includegraphics[width=\linewidth]{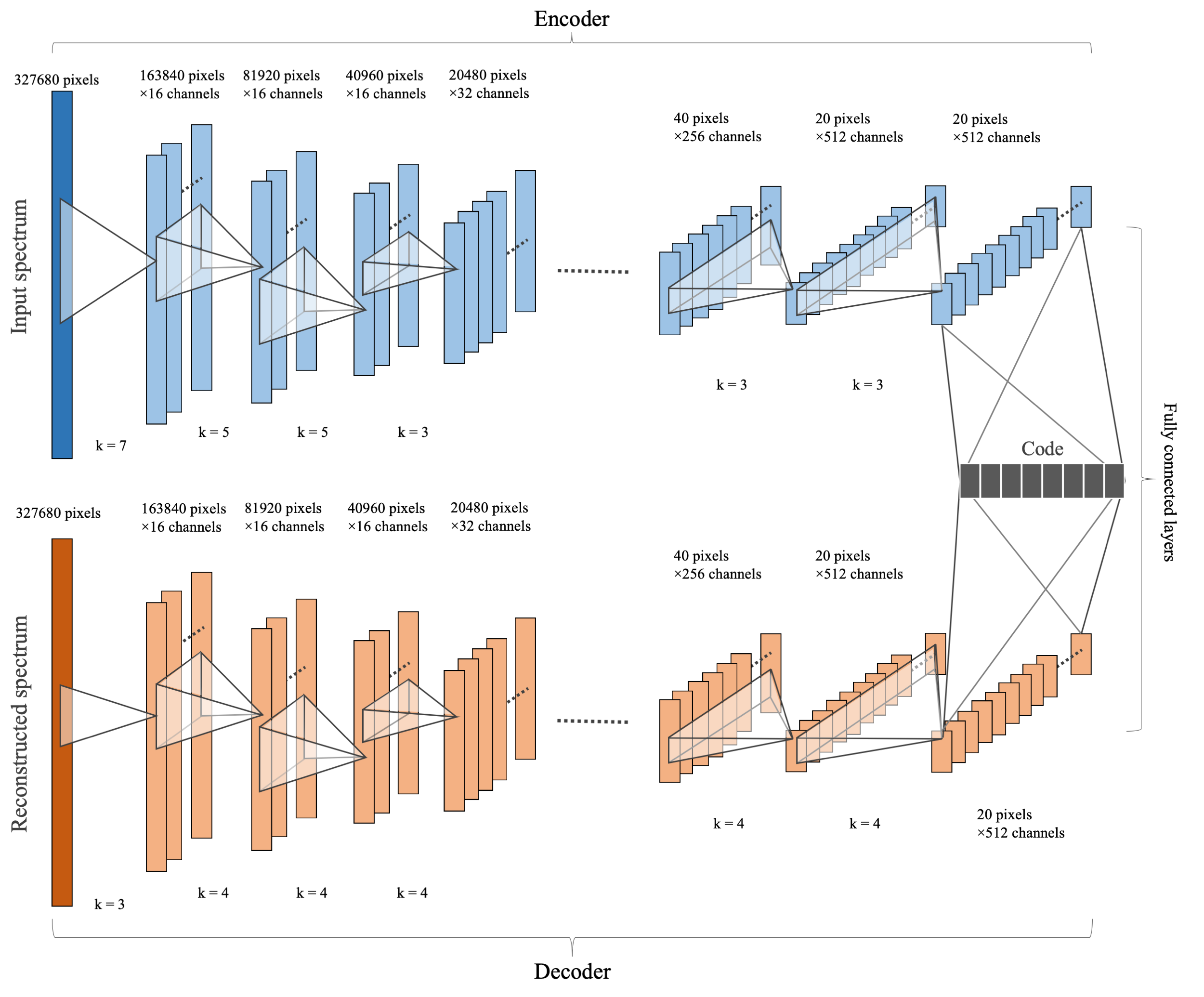}
  \end{center}
  \caption{Detailed architecture of the deterministic autoencoder. Due to lack of space, not all the layers have been visualized. The missing information can be extracted from the released source code.}
  \label{fig:architecture_detailed}
\end{figure*}

\section{Network Architecture}
\label{app:architecture}
\Cref{fig:architecture_detailed} illustrates details of the deterministic autoencoder. The VAE version follows the exact same architecture, and differs only around the bottleneck, as illustrated in \cref{fig:architecture}.

\section{Retrieving Validation Labels}
\label{app:Retrieve}

To collect a large number of existing physical labels from the literature, we use both SIMBAD \citep{wenger2000simbad} and the TESS Input Catalogue \citep[TIC;][]{stassun2019tic}, a ``compiled
catalogue of stellar parameters''.

\paragraph{Cross-Matching Process}
We produce metadata of our HARPS subset as a reference table. Each row of the table contains information regarding a single spectrum, including its position on the sky.
The table contains possibly many observations of the same star, as discussed in \cref{sec:dataset}. Moreover, the position accuracy is quite low and the photometric counterpart may be located at a distance of tens of arcseconds from the input table position.

In the first step we consider each entry of the table as an independent object. We load the reference table into TOPCAT~\citep{taylor2005topcat} and perform a cross-match with the remote VizieR version of the TIC table using the CDS Xmatch service from TOPCAT.
We perform a simple cone-search cross-match returning all TIC objects in a radius of 40 arcsec around each of the input table positions. This results in more than 5 million matches (5\,610,122 exactly): the TIC being deep, we get a large number of spurious association in such a large search box.

\begin{figure*}[!h]
  \begin{center}
    \includegraphics[width=0.3\linewidth]{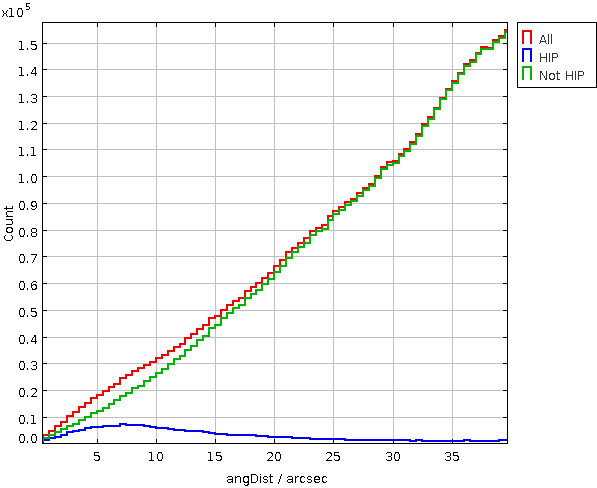}
    \includegraphics[width=0.3\linewidth]{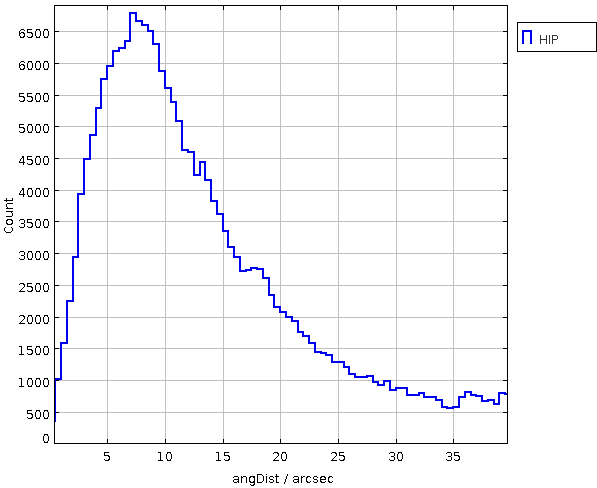}\\
    \includegraphics[width=0.3\linewidth]{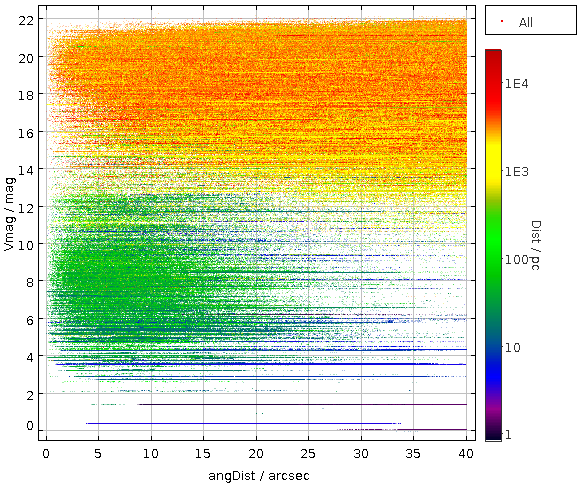}
    \includegraphics[width=0.3\linewidth]{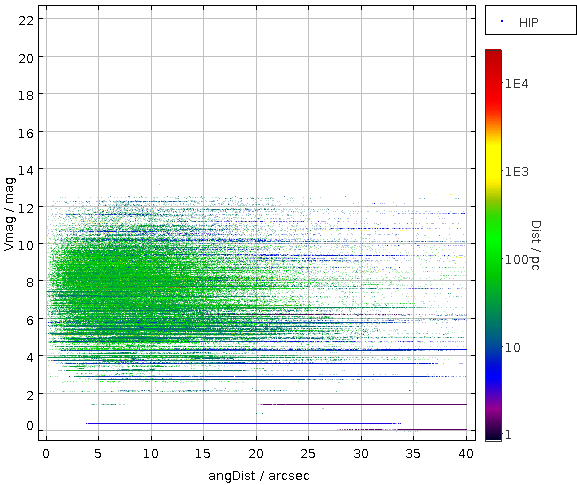}
    \includegraphics[width=0.3\linewidth]{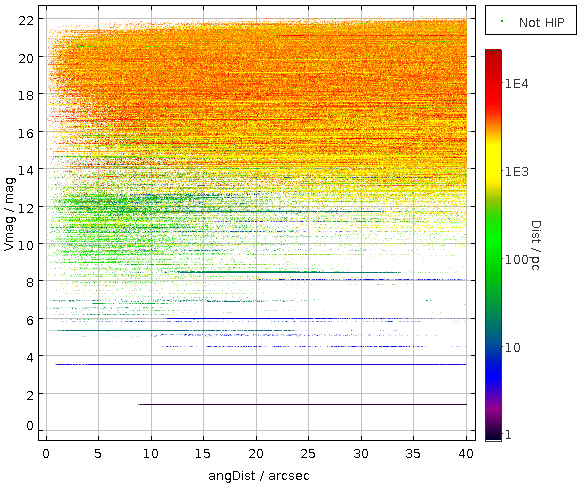}
  \end{center}
  \caption{HARPS vs TIC cross-match.
    Top left: angular separations (in arcsecond) for all cross-matches (red), the HIP selection (blue) and its complement (green).\\
    Top right: re-scaled histogram of the angular separations for the HIP selection.\\
    Bottom: angular separation (in arcsecond) versus magnitude V versus distance of the TIC stars.
    The left, center and right panels show all matches, 
    the selected HIP matches, and its complement respectively.}
  \label{fig:harps_tic_xmatch}
\end{figure*}

Plotting angular separation versus magnitude V versus distance of the TIC stars plot (\cref{fig:harps_tic_xmatch}), good matches seem to be separable from the spurious ones based on the Vmag. It complies with the prior knowledge that most HARPS objects have a magnitude lower than 12--13 (the green points on the plot).
This magnitude correspond more or less to the limiting magnitude in the Hipparcos and Tycho catalogue \citep[HIP][]{perryman1997hipparcos}. We thus decided to filter TIC data to keep only objects having an observation in HIP. With this single selection criteria, we put a (loose) constraint on magnitudes and ensure a better homogeneity of the selected sample, since all objects have been observed at least in the Hipparcos catalogue. This leaves us with 209\,183 (4\%) associations. With this selection, we probably miss good matches in a mag range of 10--12.5, the green points on the right part of the plot.

Initially, the histogram of the angular separations of all matches is dominated by spurious matches (almost only the linear -- Poisson -- component is visible). But selecting only HIP objects in the TIC catalogue, the histogram is now dominated by good matches (the linear component being quite low).

\begin{figure*}[]
  \begin{center}
    \includegraphics[width=0.3\linewidth]{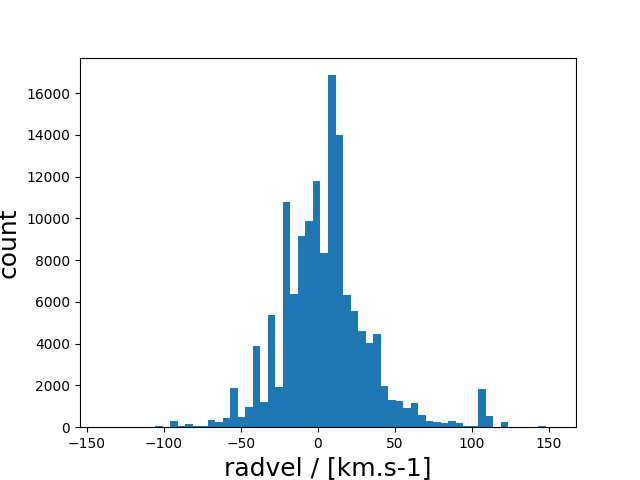}
    \includegraphics[width=0.3\linewidth]{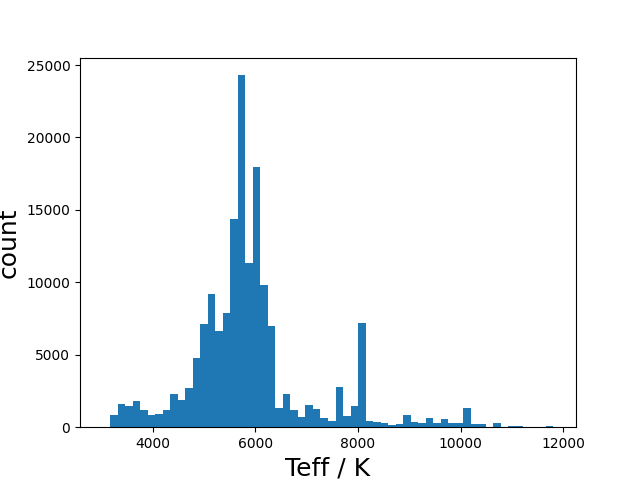}
    \includegraphics[width=0.3\linewidth]{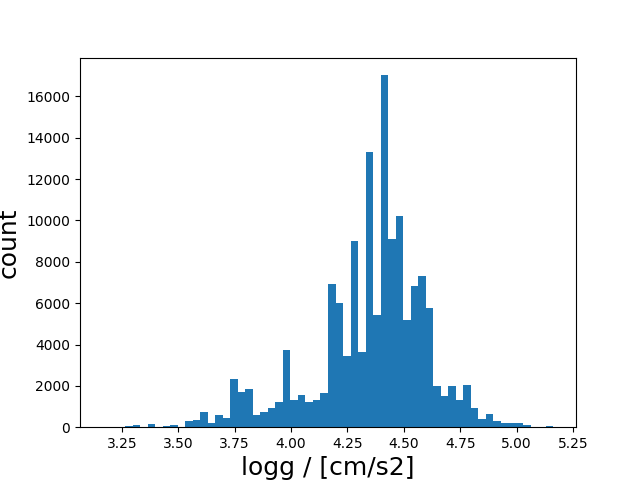}
    \includegraphics[width=0.3\linewidth]{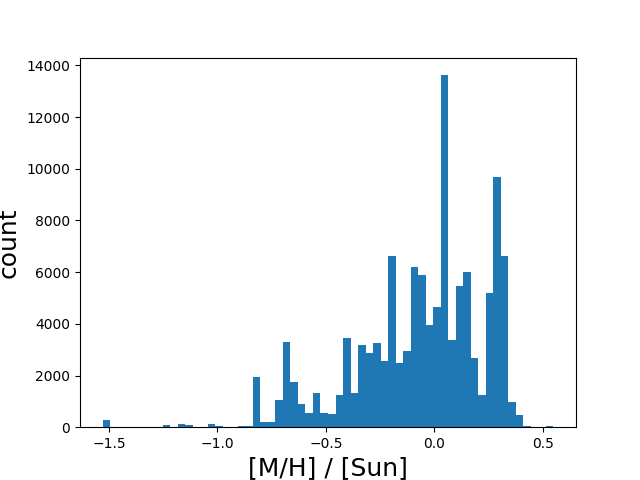}
    \includegraphics[width=0.3\linewidth]{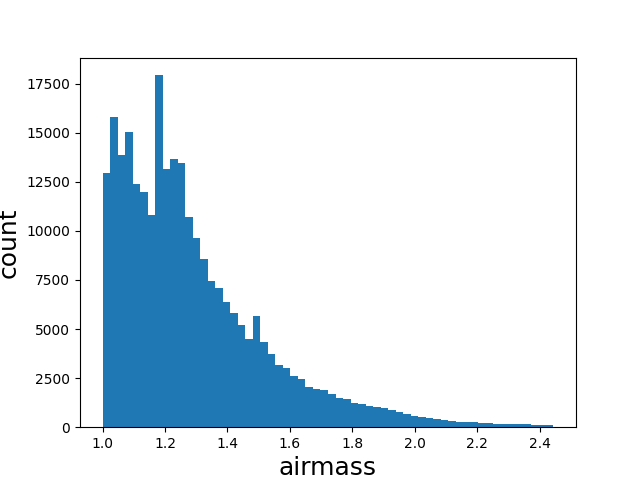}
    \includegraphics[width=0.3\linewidth]{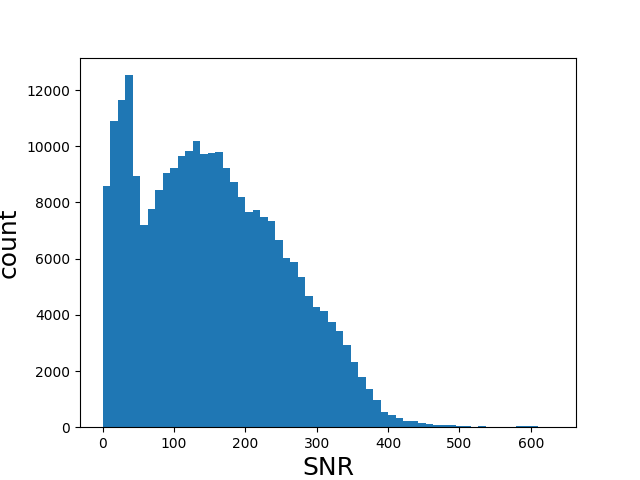}
  \end{center}
  \caption{Distributions of successfully acquired labels. It should be noted, however, that these only represent a subset of HARPS spectra, and do not necessarily represent the exact distributions of the parameters in HARPS.}
  \label{fig:label_distributions}
\end{figure*}

After this first selection, we still get HARPS entries associated to multiple TIC objects. As we favored reliability over completeness, we removed those objects resorting to an internal match in TOPCAT. We get an output 185\,662 HARPS spectra associated each with a single TIC entry.

To add SIMBAD "labels", we finally cross-match our results with SIMBAD, 
keeping the closest match in a 2 arcsec radius around the TIC object positions.
5\,000 HARPS objects are lost at this last step.

In the end, we get 179\,389 matches with
\begin{itemize}
  \item 151\,743 radial velocities
  \item 120\,440 metallicity
  \item 145\,372 mass
  \item 145\,372 log(g)
  \item 167\,728 Teff.
\end{itemize}

The cross matching and the resulting labels are by no means complete. The labels may not be quite accurate either. This, however, suffices for our validation experiments as we seek only the overall possible patterns and correlations, not the exact values.

\section{Principal Component Analysis}
\begin{figure*}[]
  \begin{center}
    \includegraphics[width=0.9\linewidth]{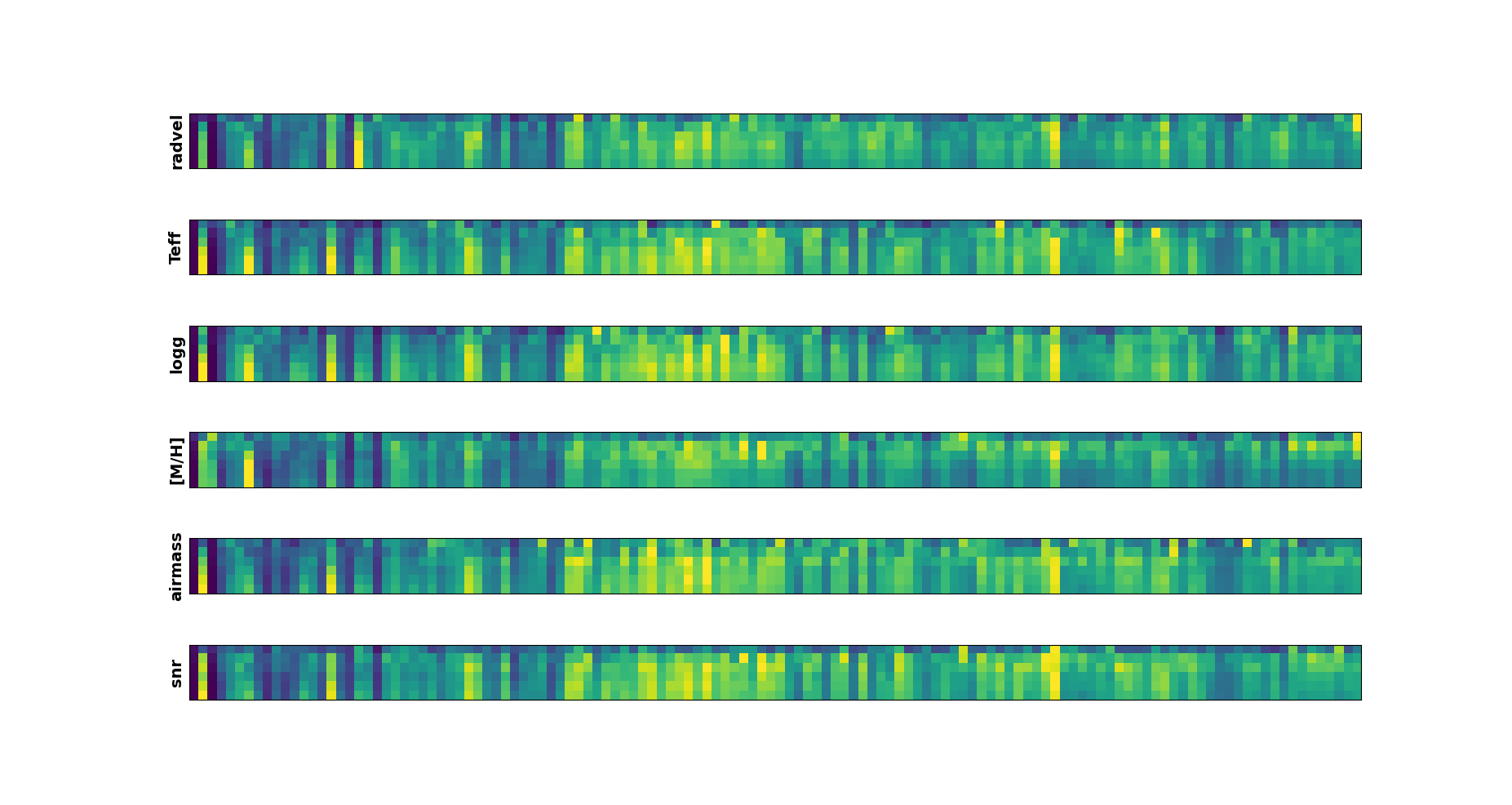}
  \end{center}
  \caption{MI indicators for detection of traces of physical parameters in the first 128 components of PCA. As expected, no clear traces of individual parameters can bee seen; in other words information about each physical parameter is spread over many dimensions.}
  \label{fig:pca}
\end{figure*}

In \cref{fig:pca} we depict how our MI indicators would work on the first 128 principal components of our dataset. As expected, being essentially a linear transformation, PCA should not be expected to result in any sort of "smart" disentanglement of features.

\label{app:pca}

\end{appendix}

\end{document}